\documentclass[aps,prb,floatfix,twocolumn,showpacs,reprint,groupedaddress]{revtex4-1}
\usepackage{float}
\usepackage{graphicx}
\usepackage{amsmath}
\usepackage{bbold}
\usepackage{color}

\begin{document}
\title{Spin-Orbit Semimetals in the Layer Groups}

\author{Benjamin J. Wieder and C. L. Kane}
\affiliation{Department of Physics and Astronomy, University of Pennsylvania,
Philadelphia, PA 19104}

\begin{abstract}
Recent interest in point and line node semimetals has led to the proposal and discovery of these phenomena in numerous systems.  Frequently, though, these nodal systems are described in terms of individual properties reliant on specific space group intricacies or band-tuning conditions.  Restricting ourselves to cases with strong spin-orbit interaction, we develop a general framework which captures existing systems and predicts new examples of nodal materials.  In many previously proposed systems, the three-dimensional nature of the space group has obscured key generalities.  Therefore, we show how within our framework one can predict and characterize a diverse set of nodal phenomena even in two-dimensional systems constructed of three-dimensional sites, known as the ``Layer Groups''.  Expanding on an existing discussion by Watanabe, Po, Vishwanath, and Zaletel of the relationship between minimal insulating filling, nonsymmorphic symmetries, and compact flat manifolds, we characterize the allowed semimetallic structures in the layer groups and draw connections to related three-dimensional systems.  
\end{abstract} 

\maketitle

\section{Introduction}
\label{sec:introduction}

Nodal semimetals are systems of arbitrary dimensionality for which the valence and conduction bands meet and form nodes in a limited set of places in the Brillouin zone.  The existence of nodal points at the Fermi energy has implications for bulk transport, surface physics, and is even related to topology~\cite{ChiralAnomaly1,ChiralAnomaly2,ChiralAnomaly3,GrapheneEdge1,GrapheneEdge2,KaneMele,3DTI}.  Since the discovery of nodal points with linear dispersion in single-sheet graphene~\cite{GrapheneDirac1,BLG1}, there has been great effort to locate and characterize similar band-touching nodes in other systems.  Rich nodal physics can also be found in graphene bilayers, however these systems have quadratically-dispersing nodes, and thus differing transport properties and gapped phases~\cite{BLG1,BLG2,BLG3,BLG4,BLG5,BLG6,BLGMine}.  Therefore, this search for graphene-like physics can be considered more specifically a search for systems with \emph{linearly-dispersing} nodal points at the Fermi energy.  

Extending this search to three dimensions, there are generically two realizations of this nodal physics: point nodes and nodal lines.  Bands can meet at nodes with linear dispersion in three directions and form Dirac or Weyl points, with four- and two-fold nodal degeneracies respectively~\cite{ChineseDirac1,ChineseDirac2,SaadDirac,NSDirac2,NSDirac3,NSDirac4,Weyl1,Weyl2,Weyl3,Weyl4,Weyl5,Weyl6,Weyl7,Weyl8}, or in unusual three-, six-, and eight-fold degeneracies~\cite{NewFermions,DDP}.  Bands can also meet and form line nodes, or lines along which there is no dispersion in one direction and linear dispersion in the remaining two directions~\cite{LineNode1,LineNode2,LineNode3,LineNode4,KeeRing,KeeExplain,LineNode7,LineNode8,LineNode9,LineNode10,LineNode11,LineNode12}.

Each of these examples of nodal phenomena owes its protection to some combination of topology and exact crystalline symmetries.  However, many of them have been described in terms of individual properties, such as topological invariants and symmetry eigenvalues.  In this manuscript, we seek to provide a more generalized consideration of those nodal systems protected by crystalline symmetries, as well as a means to sort and classify them.

\subsection{The Two Flavors of Semimetals}

As a starting point for sorting these various nodal materials, one could ask whether or not any of their nodal features can be eliminated to open a gap.  Consider the ability to remove point nodes pairwise, or to shrink and gap line nodes at a point, while still preserving a system's crystalline symmetries.  The Dirac points in Cd$_3$As$_2$ and Na$_3$Bi, as well as the line nodes  proposed in Cu$_3$N and observed in Ca$_3$P$_2$~\cite{ChineseDirac1,ChineseDirac2,LineNode1,LineNode2,LineNode3}, obey this property, whereas the proposed Dirac points in BiO$_2$ and the Dirac line nodes in SrIrO$_3$ do not~\cite{SaadDirac,KeeRing}.  We can designate this first category of nodal systems as \emph{band-inversion} semimetals.  The nodes in these systems are optional features of the space group; they are certainly locally permitted by crystalline symmetries, or topology in the case of Dirac line nodes under weak spin-orbit interaction, but they are otherwise globally extraneous.  Conversely, the nodes in BiO$_2$ and SrIrO$_3$ are part of groupings of 4 and 8 bands, respectively.  All of the bands in these systems appear at a minimum in groupings of these numbers, and the existence of nodal features at the Fermi energy appears to be guaranteed by the electron filling.  We therefore designate these as \emph{essential} semimetals, or systems with nodes which are pinned into existence by additional space-group-specific symmetries.  

This relationship between essential nodal features and filling is, as discussed in this manuscript, the single-particle manifestation of the concept of ``minimal insulating filling.'' Watanabe, Po, Vishwanath, and Zaletel (WPVZ) realized that for 220 of the 230 space groups, a discussion of related flat compact manifolds allows one to exclude the existence of an insulating state at fillings specific to each space group~\cite{WPVZ}.  Note that this criterion only addresses the allowed existence of a consistent band gap for systems in which all bands are either fully occupied or unoccupied.  It does not exclude cases where the dispersion is large and the Fermi energy cuts through electron and hole pockets.  Nevertheless, for the purposes of this manuscript, we will still refer to such indirect gap materials as insulators, as many of the analysis methods for such systems, such as polarization and topological invariant calculations, only require the ability to find well-separated band groupings.

In this manuscript, we designate the integer fillings at which WPVZ deduced an allowed insulating state as the ``WPVZ bound.''  In the limit that interactions are weak and bands are well-defined, we observe that \emph{the same combinations of crystalline symmetries which define the WPVZ bound also conspire to guarantee essential groupings of bands.}  Furthermore, in cases where local topological features might be removed to open a gap, such as the combination of two Weyl points with opposite Chern numbers~\cite{Weyl2}, this bound provides an obstruction to that process.  For example, if two Weyl points are required to exist by minimal insulating filling, they cannot be gapped out while preserving all crystalline symmetries, even if they have opposite Chern numbers, as the bands which comprise them cannot be separated without lowering the system symmetry and changing the space-group-specific filling constraints.    

In practice, the determination of this bound, as well as the accompanying analysis of crystalline symmetry algebra, can become difficult in three dimensions.  Noting that Young and Kane also predicted essential semimetallic features in two-dimensions~\cite{Steve2D}, we therefore propose a consideration of the WPVZ bound and the allowed nodal features in two dimensions.  Furthermore, to restrict our discussion to the role of crystalline symmetries, we require strong spin-orbit interaction, such that locally-protected topological features, such as the Dirac points in graphene, are disallowed.  In this paper, we observe that, considering the full set of 80 two-dimensional systems known as the ``layer groups''~\cite{LGtoSGconvert}, nontrivial WPVZ bounds can be achieved and rich nodal semimetallic phenomena can be both created in \emph{band-inversion} semimetals and required in \emph{essential} semimetals.  Some of these phenomena are protected by the same mechanisms as are their three-dimensional cousins that occur in such materials as BiO$_2$ and SrIrO$_3$.  Furthermore, we show how this analysis predicts previously uncharacterized nodal phenomena in two- and three-dimensions, such as band-inversion Dirac points protected by an inversion-center offset and an essential 8-band ``cat's cradle'' Weyl fermion feature.  

\subsection{Contents of this Manuscript}  

This paper is structured as follows.  First, in~\ref{sec:platycosms} we use a discussion of compact flat manifolds  to rederive the WPVZ bound in first purely two-dimensional systems (wallpaper groups) and then in two-dimensional systems embedded in three dimensions (layer groups).  Following that, we provide in~\ref{sec:eigenvals} a breakdown of the eigenvalue structure of band groupings as it relates to spatial symmetries and inversion centers.  Finally, in~\ref{sec:LGs} we combine both descriptions to produce criteria for predicting semimetallic features and apply them to a set of related simple models characteristic of both band-inversion and essential nodal semimetallic features in two-dimensional crystals with strong spin-orbit interaction.

\section{Platycosms and Minimal Insulating Filling}
\label{sec:platycosms}

\begin{figure}
\centering
\includegraphics[width=3.5in]{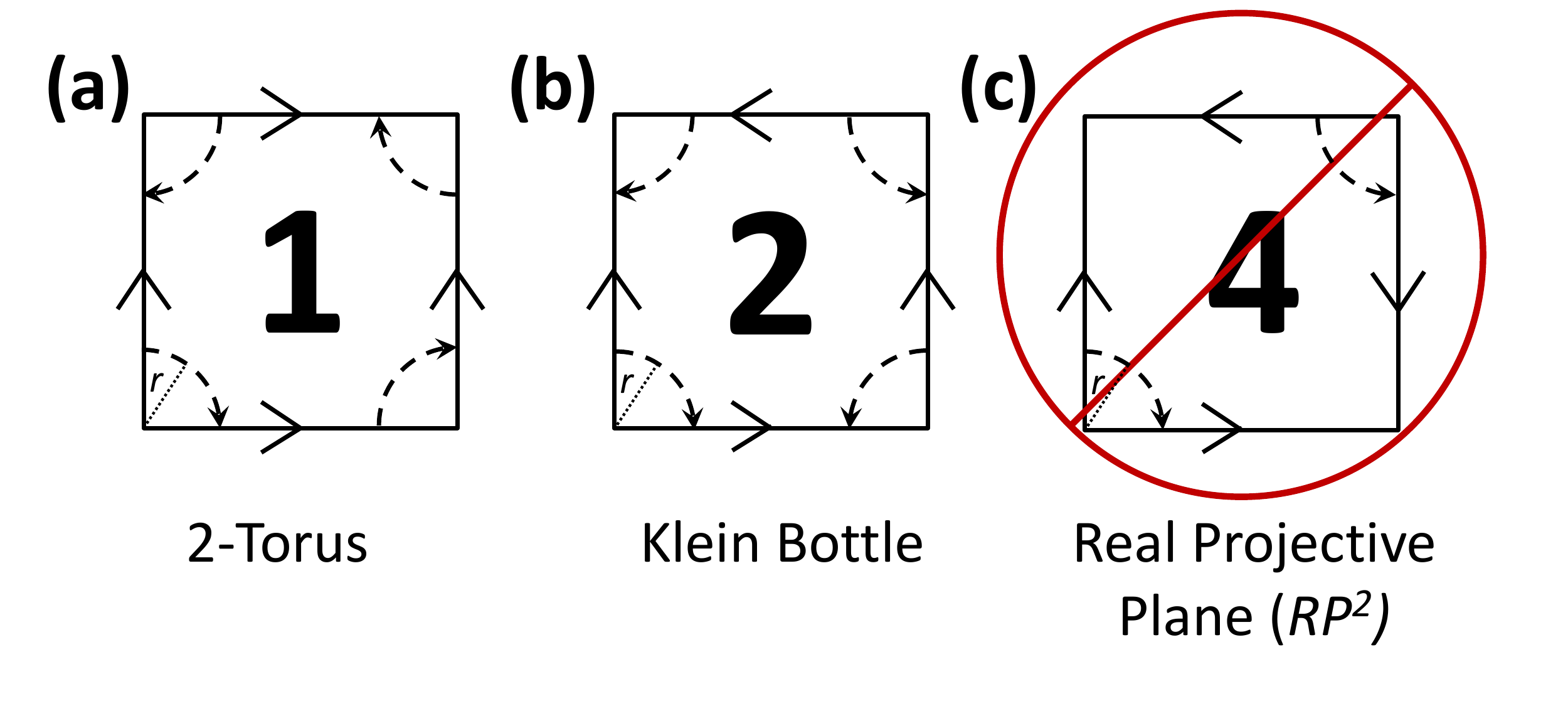}
\caption{The 3 compact manifolds which can be achieved by twisting the coordinate-axis-direction boundary conditions for a strictly two-dimensional system that is periodic in both in-plane directions.  The local designations of the perpendicular direction are indicated by the arrows, a notation known as the ``fundamental polygon.''  Of the possible manifolds, the 2-torus (a) and the Klein bottle (b) are flat, but the real projective plane ($RP^{2}$) (c) is not.  Flatness can be evaluated by testing for the existence of fixed or special points by drawing a circle centered on the boundary and comparing its circumference to that of a circle drawn on the interior.  Staring at the bottom left corner, the dashed line indicates the boundary of a circle of radius $r$.  For the 2-torus and the Klein bottle, this boundary explores all four corners, resulting in a circumference of $2\pi r$, matching the value on the interior.  However, for $RP^{2}$, this boundary only additionally explores the top right corner before returning, resulting in a reduced circumference of just $\pi r$, and indicating that the two bottom corners are special points, distinct from the interior and from each other, and therefore that $RP^{2}$ is not uniform.  The bold numbers indicate the number of times that each pattern would have to be repeated to create a supercell with the same boundary conditions as the initial 2-torus.  Pictorial guides to forming such supercells can be found in Appendix~\ref{appendix:manifolds}.}
\label{FigWallPoly}
\end{figure}

To begin, we relate the idea of minimal insulating filling to the compatibility between a given two-dimensional group and a set of flat compact manifolds.  Recent work by Watanabe, Po, Vishwanath, and Zaletel (WPVZ) has revived interest in this relationship between the space groups and three-dimensional manifolds~\cite{WPVZ}.  To introduce this topic, we begin with a review of familiar manifolds in two dimensions.  We then extend this discussion to encompass layered three-dimensional systems.  Finally, we conclude this section with a review of the WPVZ arguments for minimal insulating filling and how those arguments relate to the sets of compact flat manifolds in two and three dimensions.  

This discussion is intended as an introduction to topics in topology and group theory such as compact manifolds and group modding procedures.  At the level of this manuscript, we loosely equate such concepts as flatness, uniformity, and the absence of fixed points.  For a more formal treatment of this material, we recommend that one consult Conway and Rossetti~\cite{Conway}.  

\subsection{Compact Flat Manifolds in Two and Three Dimensions}
\label{sec:polygons}

For a 2D system periodic in one of the in-plane directions, say a piece of paper wrapped onto itself, there are two ways to assign boundary conditions.  Linking the two opposite sides without any twists produces a cylinder, and twisting the paper once produces the familiar M{\"o}bius strip.  We can also consider a 2D object with periodicity in both of the in-plane directions.  Visually, one can assign boundary conditions for such an object by considering the orientation of arrows on the fundamental polygon (Fig.~\ref{FigWallPoly}).  

The three distinct combinations of arrow assignments for a two-dimensional system with two periodic directions give a 2-torus, a Klein bottle, and the real projective plane ($RP^{2}$).  Keeping the arrow assignments consistent relative to the left-hand side of the polygon, one could produce a pattern like the one on a 2-Torus (Fig.~\ref{FigWallPoly}(a)) by linking together \emph{twice} a Klein bottle (Fig.~\ref{FigWallPoly}(b)) or \emph{four times} $RP^{2}$ (Fig.~\ref{FigWallPoly}(c)).  The creation of such a supercell from the various manifolds is visually depicted in Appendix~\ref{appendix:manifolds}.  

These manifolds can accommodate the set of truly two-dimensional systems, that is, systems comprised of two-dimensional objects embedded in a two-dimensional space.  Some of these manifolds permit the embedding of a subset of systems known as the \emph{Wallpaper Groups}~\cite{WallpaperConway}.  When considered in three dimensions, the wallpaper groups consist of the 17 unique combinations of symmetries which could describe the two-dimensional boundary of a three-dimensional object.  They are therefore characterized only by symmetry operations which preserve the interior and exterior of such a boundary, namely mirror and glide lines in the plane and rotations of the surface about its normal vector.  For this reason, the operation of an arrow flip could also be considered the modding out of a glide mirror, as a site on the left side of a \emph{flipped} boundary is related to a site on the right side by a glide mirror operation.  

Modding out a single glide produces the Klein bottle (Fig.~\ref{FigWallPoly}(b)): a two-dimensional manifold that has just one surface (also sometimes called nonorientable as its surface does not preserve the handedness of a coordinate axis which traverses it)~\cite{Conway}.  This object is compact, as it was constructed periodically, and but it is also flat. Modding out an additional glide produces the one-sided real projective plane ($RP^{2}$) (Fig.~\ref{FigWallPoly}(c)), which unlike the 2-torus and the Klein bottle has fixed points, and is therefore not flat and uniform.  

One can test for these fixed or special points in two dimensions by considering the circumference of a circle of radius $r$ centered throughout the manifold.  In the center of all three  two-dimensional manifolds, this circle has circumference $2\pi r$.  The only possible deviations from this value can occur at the boundaries of the manifolds, as depicted by the dashed lines in Figure~\ref{FigWallPoly}.  For the 2-torus and the Klein bottle, the boundary of a circle of radius $r$ centered at the bottom left corner still explores all three other corners before returning, resulting in a circumference of $2\pi r$, a value that matches circles on the interior.  However, for $RP^{2}$, the boundary returns to itself after only reaching the top right corner, giving a circumference of just $\pi r$, and indicating that the two bottom corners are special points, different from each other and from the interior.  

A key restriction of WPVZ's argument is that any manifold for consideration must allow the uniform embedding of a crystal lattice such that the Bloch wave functions are periodic plane waves~\cite{WPVZ}.  The special corners in $RP^{2}$ violate this uniformity, and therefore modding out onto it is disallowed.  We therefore conclude that \emph{the wallpaper groups can only be uniformly embedded on the 2-torus and on the Klein bottle.}  

More formally, we can consider this restriction to fixed-point-free manifolds in terms of group theory.  WPVZ take advantage of work by Bieberbach that established that a collection of 10 space groups are fixed-point free.  These space groups each have only one nonsymmorphic symmetry, or have a very special combination of two perpendicular nonsymmorphic symmetries such that no fixed points are introduced.  We note that for just one of these 10 space groups, there is a corresponding wallpaper group. This wallpaper group, $pg$ (space group 7 $Pb11$), is characterized just by a single glide symmetry.  The process of modding out a glide and flipping an arrow on the fundamental polygon is more formally asking whether or not $pg$ is a subgroup of that wallpaper group, such that it can be modded out.  As $pg$ only contains one glide, modding by it results in a placement onto the Klein bottle, and as none of the remaining nonsymmorphic wallpaper groups correspond to fixed-point-free space groups when stacked in three dimensions, the Klein bottle is the \emph{only} manifold besides the 2-Torus which is compatible with the embedding of a purely two-dimensional crystal lattice.  Finally, as only nonsymmorphic operations can provide a coordinate-arrow twist of the fundamental polygon, and as all of the nonsymmorphic wallpaper groups describe rectangular lattices, we therefore find that \emph{all} possible restrictions on the uniform embeddings of 2D lattices can be obtained by considering whether or not axis boundary condition flips on the fundamental polygon introduce fixed points.     

\begin{figure}
\centering
\includegraphics[width=3.5in]{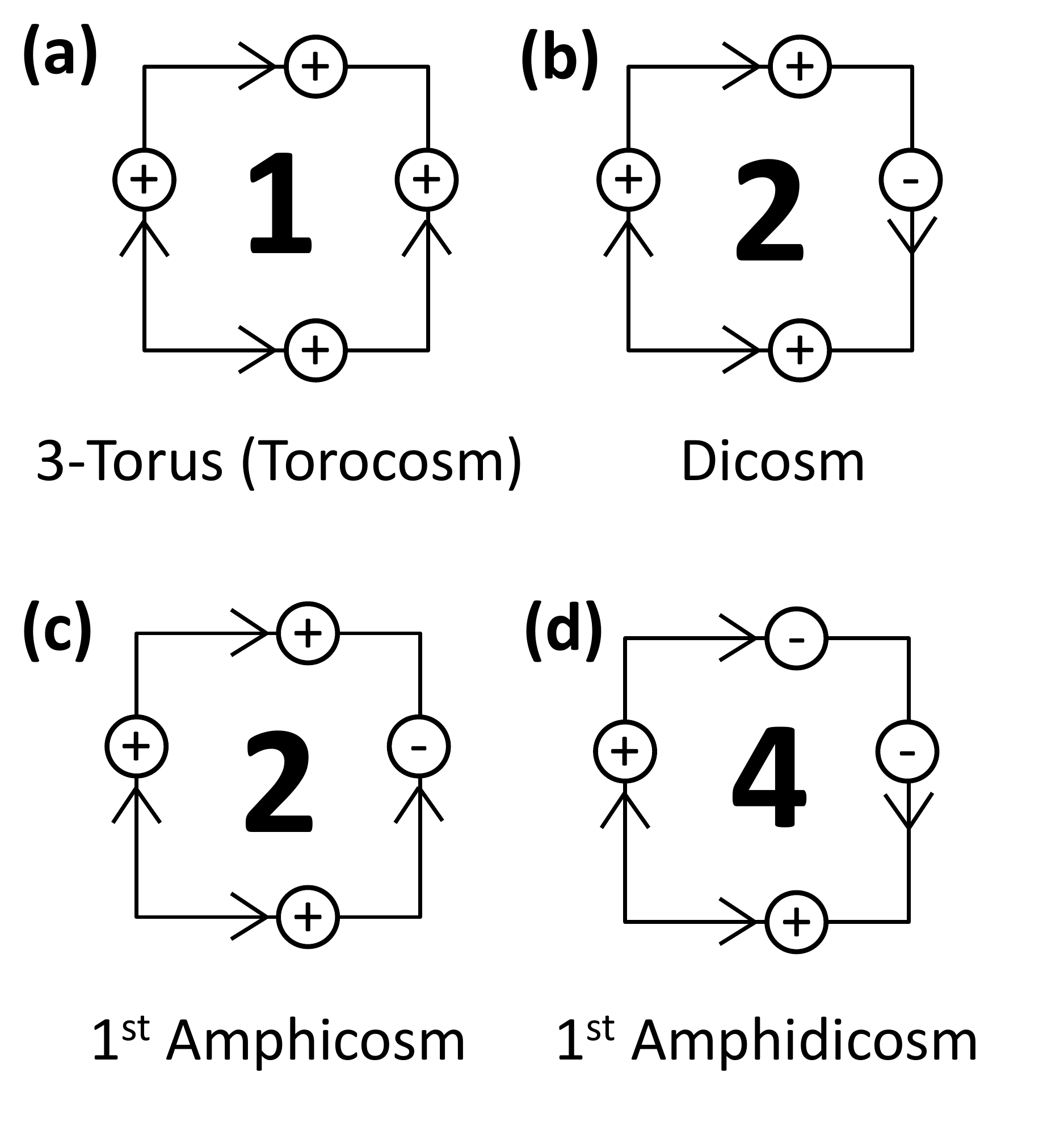}
\caption{The compact, flat manifolds which can be achieved in three-dimensional, layered systems.  The notation used is a modification of the fundamental polygon from Figure~\ref{FigWallPoly} by the additional local assignment of the stacking direction as indicated by the $\oplus$ and $\ominus$ signs.  Unlike in the strictly two-dimensional wallpaper systems, the modding out of a two-fold screw is also allowed and leads to a new manifold which doesn't decompose into $S^{1}$ multiplied by a wallpaper manifold.  Modding out a screw preserves the interior and exterior surfaces leading to the dicosm (b).  Modding out a glide reduces the system to being one-sided and leads to the 1st amphicosm (c).  In these layered systems, unlike in the 2D wallpaper cases, a particular combination of two perpendicular nonsymmorphic operations can be modded out without introducing fixed points, leading to a new flat manifold: the 1st amphidicosm (d).  The bold numbers indicate the number of times that each pattern would have to be repeated to create a supercell with the same boundary conditions as the initial 3-torus (a).  The procedure for forming such supercells is explained with visuals in Appendix~\ref{appendix:manifolds}.}
\label{FigLayerPoly}
\end{figure}

In three-dimensions, this argument can be extended, though unfortunately in general there is no neat analogue to the fundamental polygon to describe the equivalent operations of modding out glide mirrors or screws.  The 10 resultant three-dimensional flat manifolds are known as the platycosms.  Descriptions of them, as well as, where possible, connections to more familiar manifolds, are detailed in plain language in Ref.~\onlinecite{Conway}.  

However, for 80 realizations of 70 three-dimensional space groups, crystals can be decomposed into two-dimensional sheets, only related to the next layer by lattice periodicity in the stacking direction.  For these layered systems, symmetry-enforced physics is entirely determined by the symmetries of a two-dimensional single layer, and therefore even though they are three-dimensional systems, many of their properties can be determined by confining analysis to the two-dimensional subsystem of a single layer.  These single-layer systems, comprised of two-dimensional objects embedded in three dimensions, are a subset of ``subperiodic groups'' known as the \emph{Layer Groups}~\cite{LGtoSGconvert}. 

For these layer group systems, placement onto a platycosm can be visually represented by modifying the fundamental polygon to include a local specification of the $\hat{z}$ direction (stacking direction) (Fig.~\ref{FigLayerPoly}).  For these systems the modding out of a two-fold screw is now also permitted in addition to the glide mirror mod, with the screw mod taking one to the two-sided dicosm (Fig.~\ref{FigLayerPoly}(b)) and the glide mod taking one to the one-sided 1st amphicosm (Fig.~\ref{FigLayerPoly}(c)) (which can also be expressed as $klein\ bottle\times S^{1}$)~\cite{WPVZ,Conway}.  Remarkably, the modding out of a glide which flips the $\hat{z}$ direction as well as a perpendicular screw results in a manifold which, unlike $RP^{2}$ in the fully-two-dimensional wallpaper systems, is flat.  This manifold, the 1st amphidicosm (Fig.~\ref{FigLayerPoly}(d)), actually needs to be placed \emph{four} times to reorient the system boundary to match the torus configuration of the modified fundamental polygon, a property which will have significant band-structure implications.  Visual representations of this comparison of the twisted manifolds with the initial torus can be found in Appendix~\ref{appendix:manifolds}.  

\subsection{Minimal Insulating Filling by Kramers' Theorem}
\label{sec:minfill} 

In Ref.~\onlinecite{WPVZ}, the authors combined an understanding of the platycosms with Kramers' theorem to make a strong statement about space-group-symmetry-enforced obstructions to insulators occurring at specific fillings.  In the following text, we reproduce their arguments and then apply them to the restricted set of layer group systems.  

For any real system with spinful electrons, there exists a time-reversal operator $\theta$ which squares to $-1$ and mandates, according to Kramers' theorem, that each state is two-fold-degenerate.  Under this restriction, a periodic crystal with an odd number of electrons $N_{e}$ must, independent of how its boundary conditions are applied, have a partially filled state and therefore be a metal or a semimetal~\cite{WPVZ}.  That number of electrons can be expressed as $N_{e} = N_{cell}\nu$, where $N_{cell}$ is the number of unit cells and $\nu$ is the filling per unit cell.  Therefore, Kramers' theorem can be restated as a requirement that $\nu\in 2\mathbb{Z}$ to avoid a metallic state.  

\begin{figure}
\centering
\includegraphics[width=3.5in]{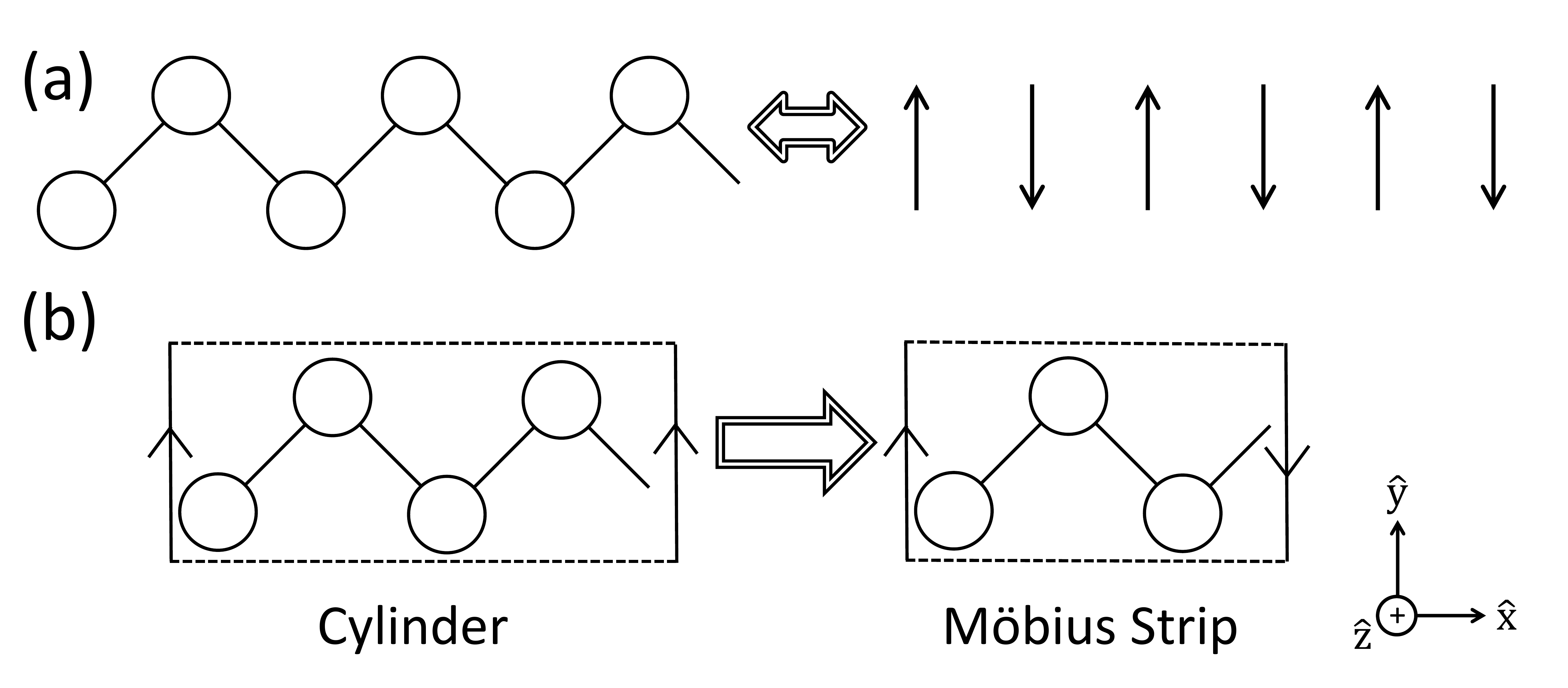}
\caption{The Su-Schrieffer-Heeger (SSH) Model, tuned to its quantum critical point, on a periodic system.  In this limit, the two sublattices are energetically identical, and are therefore related by the nonsymmorphic operation of a mirror of the $y$-direction followed by a half-lattice translation in the $x$-direction.  All of the information about this system can be encoded by mandating that both sublattices live at the same potential and by replacing each site with a vector object pointing in the $\pm\hat{y}$ direction (a).  When the system has an even number of sites and the boundary condition is not twisted, it lives on a cylinder (b).  However, one can produce an electronically equivalent system, if the number of unit cells is large, by removing one site and twisting the axis boundary condition on the $y$ direction, which places the lattice instead on a M{\"o}bius strip.}
\label{FigSSH}
\end{figure}

Now, when the unit cell is constructed of energetically identical sites related by some spatial operation, one can consider the operation which takes you between sites as a nonsymmorphic symmetry: a combination of a fractional lattice translation and a spatial operation.  The quintessential example of a system with such a symmetry is the undimerized Su-Schrieffer-Heeger (SSH) model, which is a one-dimensional bipartite chain where the sites can be equivalently represented by a local up or down vector object (Fig.~\ref{FigSSH}(a)).  While the gapped phases of this model already have a symmorphic mirror symmetry in the $x$ direction, its quantum critical point, reached by enforcing the same chemical potential on each sublattice, possess a second, \emph{nonsymmorphic} symmetry, one which relates the two sublattices by a mirror of the $y$ direction and a half-lattice translation in the $x$ direction.   

In this limit of the SSH model, one could consider taking the final unit cell before the boundary and eliminating one of the sites.  This new system with a decimated unit cell, if sufficiently large, will still be electronically equivalent to the original system as long as the boundary condition reproduces the nonsymmorphic symmetry (Fig.~\ref{FigSSH}(b))~\cite{WPVZ}.  However, the resultant crystal now has a fractional $N_{cell}$ (here a half integer), and therefore in this crystal, assembled with this boundary condition, Kramers' theorem actually requires that $\nu\in 4\mathbb{Z}$ to avoid a metallic state.  Stated differently, \emph{four-band models of this SSH model tuned to have a nonsymmorphic symmetry are obstructed from being bulk insulators at half filling}.  One can consider the SSH quantum critical point as a manifestation of that obstruction.

In systems with more than one periodic direction, there can be multiple nonsymmorphic symmetries to consider.  The process of decimating the final unit cell and twisting the axis boundary condition to reproduce the nonsymmorphic symmetries is in fact just a reproduction of the modding process used to generate the manifolds in Section~\ref{sec:polygons}.  In 1D (with two-dimensional sites) with this SSH model, the decimation process reduces the underlying manifold from a cylinder to a M{\"o}bius strip~\cite{Frieze}.  In two and layered three dimensions, this decimation process exactly corresponds to placing the lattice onto one of the manifolds in Figures~\ref{FigWallPoly} and~\ref{FigLayerPoly} respectively.  With multiple nonsymmorphic symmetries, \emph{the condition for flatness is a simple check in these systems whether the product of the two symmetry operations modded out, as defined from a common origin, is itself also an inherently nonsymmorphic operation.}  The prefactor on the integer filling to avoid a metallic state exactly corresponds to the degree of decimation of the final unit cell, expressed as the large bold numbers in Figures~\ref{FigWallPoly} and~\ref{FigLayerPoly}.  For reference, visual representations of this decimation process for two-dimensional lattices can be found in Appendix~\ref{appendix:manifolds}.  

In the limit that interactions are weak and bands are well defined, \emph{this minimum filling corresponds to the minimum number of bands which must be tangled together, independent of any band-tuning conditions}.  These essential tangles of bands are bounded by two- and four-fold degeneracies that can only be moved, but not gapped, by tuning space-group-allowed hopping terms.  Any band crossings on lines between these degeneracies are therefore also stuck in existence, and are also only capable of movement but not gapping.  Throughout this manuscript, we will relate these groupings of bands to the particular WPVZ bounds for minimal insulating filling established by combining the platycosm modding procedure with Kramers' theorem.  

Finally, WPVZ noted that in three dimensions, there are a handful of exceptions to the platycosm formulation of minimal insulating filling established by these arguments, with breakdowns occurring in some cases of multiple nonsymmorphic symmetries or because of unusual, highly degenerate points~\cite{NewFermions,DDP}.  However, none of these exceptions occur in the 80 layer groups, and therefore the platycosm formulation of the WPVZ bound in these systems exactly captures the filling restrictions imposed by nonsymmorphic symmetries.   

In Appendix~\ref{appendix:fillingcond}, we list all of the layer groups as well as their corresponding space groups, allowed manifold placements, and insulating filling restrictions.  

\section{Band Multiplicity and Eigenvalue Structure in the Layer Groups}
\label{sec:eigenvals}

In the band theory limit, the arguments made by WPVZ must still be consistent with any symmetry- or topology-related mechanism for the protection of nodal features.  In three-dimensional systems with strong spin-orbit coupling, nodes such as Weyl points can be locally protected by a topological invariant~\cite{Weyl2}.  However, in two dimensions with strong spin-orbit interaction, or in three-dimensional systems with a higher symmetry, the protection of nodal features is determined instead by the local symmetry eigenvalue structure of the bands.  In particular, to protect a node in strong spin-orbit systems in two dimensions, the bands which cross must not share the same symmetry eigenvalues of all simultaneously compatible (\emph{i.e.} commuting) symmetry operations, or in general they will anticross and form a gap.  

The set of protected degeneracies in the layer groups with strong spin-orbit interaction, including both essential- and band-inversion-type nodes, is therefore entirely determined by the kinds of allowed band multiplicities and eigenvalue structures.  In this section, we review basics regarding the treatment of symmetry operators in $k$-space, working up to how two-fold symmetries can provide more exotic degeneracies and eigenvalue pairings.  

In $k$-space, we can consider an operation for symmetry evaluation if the rotations, inversions,  and time-reverses in it return $k$ to itself modulo $2\pi$.  At a generic, low-symmetry value of $k$, only the combination of $P$ and $\theta$ can be a symmetry.  If $\theta^2=-1$, then $\tilde{\theta}^{2} = (P\times\theta)^{2} = -1$ enforces Kramers' theorem for each value of $k$ such that bands everywhere are two-fold-degenerate. 

Other symmetries are valid along points, lines, and planes and can also lead to two- or even four-fold band multiplets.  In this section, we examine the examples of band multiplicity and symmetry eigenvalue character which can locally protect a band crossing in layer group systems.  We start at the time-reversal-invariant momenta and reduce symmetry from there to lines and planes.  We close with a discussion of symmetry eigenvalues, working up from singly-degenerate bands to eigenvalue structures in band multiplets.

\subsection{Time-Reversal-Invariant Momenta}
\label{sec:TRIMs}

At a Time-Reversal-Invariant crystal Momentum (TRIM), the layer groups will host symmetry-required degeneracies of 2 or 4 bands for systems with time-reversal-symmetry $\theta$.  Kramers' theorem requires that under $\theta^2=-1$, states at the TRIMs are two-fold-degenerate.  Additionally, should $\theta|u\rangle\neq\Pi|u\rangle\neq|u\rangle$ , where $\Pi$ is an arbitrary symmetry operation valid at that particular TRIM, the Hamiltonian must have a degeneracy of at least 4.  A common example of this relationship between $\theta$ and $\Pi$ occurs with two-fold symmetries, for which \emph{if two spatial operations have representations which commute with time-reversal and anticommute with each other, and at least one of them squares to $+1$, then states at that TRIM will be 4-fold-degenerate}.  Four-fold degeneracies can also occur at points owing to the relationship between a spatial symmetry and a rotation of order $n>2$~\cite{ChineseDirac1,ChineseDirac2}, but in the layer groups, these systems are unable to host many of the \emph{essential nonsymmorphic} nodal features on which this manuscript focuses, and therefore we will restrict our discussion to systems with two-fold rotations. 

\subsection{Crystalline Symmetries} 
\label{sec:nonsym}

Away from the TRIMs, bands along lines and planes can be eigenstates of rotation, mirror, glide mirror, and screw rotation.  The eigenvalues of these operations are independent of their position-space origins, though their relative commutation relations, as we will see, are not.  Spatial inversion $P$ ($\vec{k}\rightarrow\vec{-k})$, valid only at the TRIMs, does not involve the spin degree of freedom, and thus independent of the square of time-reversal always has eigenvalues $\pm 1$.  For the remaining operations, we will restrict ourselves to the case where $\theta^{2}=-1$.  Rotations about an axis ($C_{n\vec{v}}$ where the rotation is through an angle $2\pi/n$ about $\vec{v}$) have eigenvalues $(-1)^{1/n}$ and are valid along high-symmetry lines in 2D and 3D.  Mirror, or an improper rotation, can be considered the product of $P$ and $C_{2\vec{v}}$ and therefore has eigenvalues $\pm i$.  Mirrors are valid along planes in 3D and lines in the layer groups, except for $M_{\vec{z}}$ which is valid for the whole 2D BZ.  

A nonsymmorphic operation, a glide or a screw, can be considered as a mirror or a rotation $g$ about some point in space, followed by a fractional lattice translation $\vec{t}$ in a direction such that $g\vec{t}=\vec{t}$.  They are valid along the same BZ lines and planes as are their symmorphic counterparts.  For a half translation, these operations (rotation or mirror) take on the same eigenvalues as their symmorphic counterparts, $\pm i$, multiplied by $e^{i\vec{k}\cdot\vec{t}}$:

\begin{equation}
\lambda^{\pm}_{2-fold\ NS} = \pm ie^{i\vec{k}\cdot\vec{t}},\ \vec{t}\cdot\vec{G}=\pi
\label{2foldNS}
\end{equation}

where $\vec{G}$ is a reciprocal lattice vector such that $\vec{t}$ is a half-lattice translation.  In the layer groups, the consideration of nonsymmorphic symmetries is greatly simplified as there are \emph{only} two-fold screws and glide mirrors; higher-fold screws or quarter-translation ``$d$'' glides require some amount of translation or rotation into the stacking or $z$ direction.  At the representation level, nonsymmorphic symmetries always have the same square at $\Gamma$ as their symmorphic counterparts, a value which winds by $-1$ as one moves along the half-translation direction in the BZ. 

As highlighted in Ref.~\onlinecite{Steve2D}, the fractional translation of a nonsymmorphic symmetry gives the nonsymmorphic eigenvalues a $k$ periodicity greater than the $2\pi$ of the Brillouin zone.  For a half translation, the eigenvalues wind with a $4\pi$ periodicity, which dictates that the $+$ and $-$ eigenstates of a glide or a two-fold screw have to connect somewhere along the translation direction and resolve this discrepancy.  In fact, this implies that the choice of designating a band as a $+$ or $-$ eigenstate away from $\Gamma$ is a gauge choice, as the eigenvalues cannot be defined continuously and with period $2\pi$.  
	
This inability to define a $2\pi$-periodic gauge for nonsymmorphic symmetries, and the resolution that bands are required to cross, is in fact the band-theory limit of the WPVZ bound on the minimal insulating filling.  For all two-fold nonsymmorphic systems, bands can be found in groupings of \emph{no fewer than 4}, and so at fillings other than $4\mathbb{Z}$, these systems are unavoidably metallic.  Furthermore, even when nodal features at fillings $\nu\neq 4\mathbb{Z}$ have well-defined topological indexes, such as Chern numbers for Weyl points in three dimensions, this requirement that 4 or more bands must be tangled together in two-fold nonsymmorphic systems obstructs two nodes from combining and gapping out.  As we will see in the examples throughout this manuscript, \emph{even if two Weyl points have opposite Chern numbers, they may not be pairwise eliminated if they are part of a nonsymmorphic-symmetry-enforced tangle of bands}.  

\begin{figure}[!]
\centering
\includegraphics[width=3.5in]{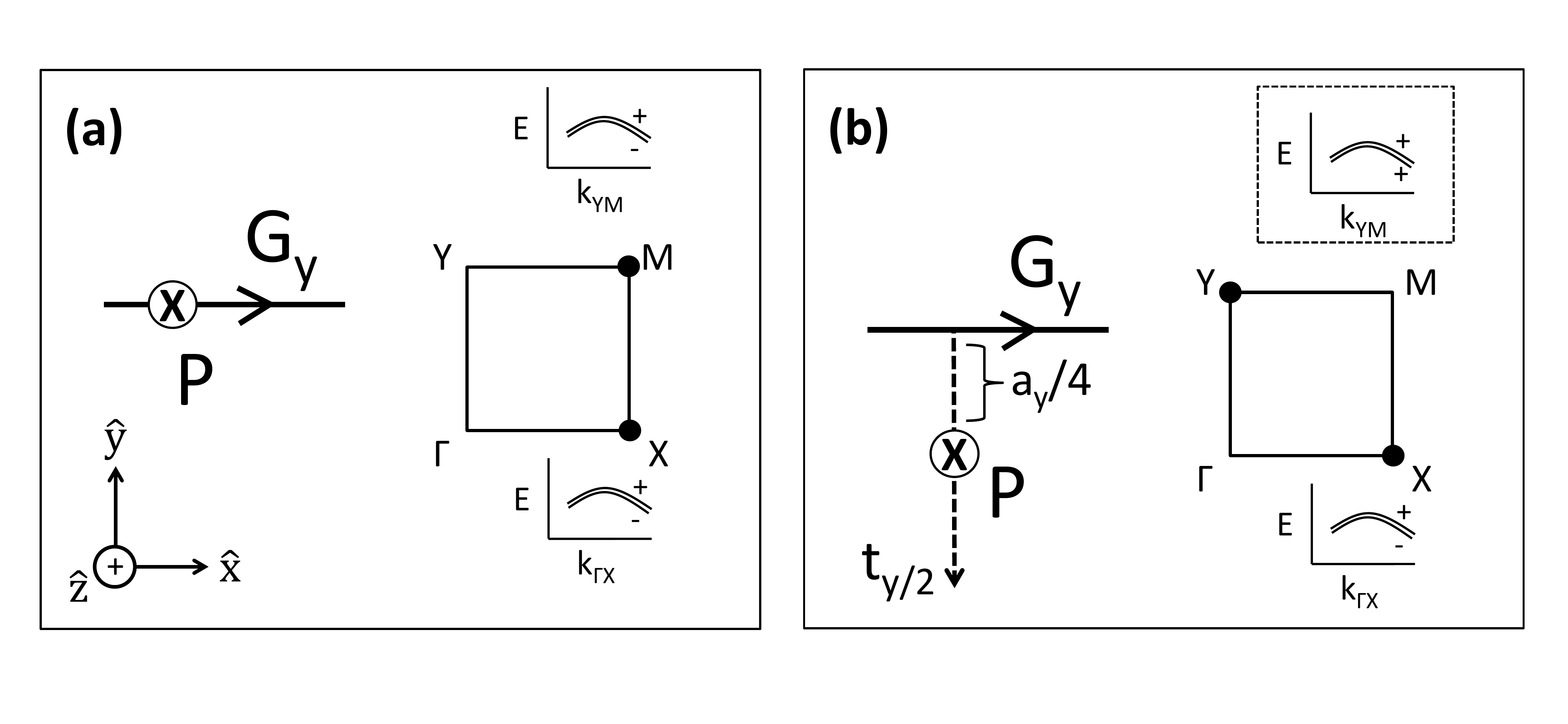}
\caption{Possible locations of a glide line $G_{y}$ relative to an inversion center $P$ ($\otimes$) in a 2D rectangular system with TRIMs $\Gamma X M Y$.  In both cases, for $\theta^{2}=-1$ all states are two-fold-degenerate because there is a local time-reversal operator $(P\theta)^{2}=-1$.  If the inversion center is coincident with the glide line (a), there will be four-fold representations at $X$ and $M$, but all eigenstates of $G_{y}$ will have eigenvalue pairings $\{+,-\}$ and can never cross.  If the inversion center differs from the glide line by a quarter-lattice spacing $a_{y}/4$ (b), then the operator for $G_{y}$ will contain an extra $t_{y/2}$ when defined from the common origin of the inversion center, leading to four-fold points instead at $X$ and $Y$.  In this case, then while the bands along $\overline{\Gamma X}$ are still paired with $G_{y}$ eigenvalues $\{+,-\}$, bands along $\overline{YM}$, if two-fold-degenerate, will be characterized by $G_{y}$ eigenvalue pairings $\{+,+\}$ or $\{-,-\}$ and can cross and create four-fold Dirac points with local protection~\cite{ChenPT}.}
\label{FigPT}
\end{figure}

\subsection{Two- and Four-Fold-Degenerate Band Multiplets}
\label{sec:multiplets}

If bands are singly degenerate, then the determination of their symmetry eigenvalues for all simultaneously compatible symmetry operations is sufficient for 2D systems with strong spin-orbit interaction in determining if bands can cross and form locally-protected nodal features.  However, when bands are two- or four-fold-degenerate, then one additionally has to determine the symmetry eigenvalues of all bands in the multiplet.  Absent the consideration of rotations of order $n>2$, a two-fold-degenerate state can occur at a point if either the representations of two spatial operations valid  at that point anticommute, or if the combination of a spatial operation $\Pi$ and $\theta$ return $k$ to itself at that point \emph{and} $(\Pi\times\theta)^{2}=-1$, locally enforcing Kramers' theorem.  In the layer groups, two kinds of spatial operations $\Pi$ support these conditions: spatial inversion $P$ and two-fold nonsymmorphic symmetries.  

When the center of inversion $P$ lies on all mirror and rotation lines, this picture is greatly simplified.  Taking $G_{y} = t_{x/2}M_{\hat{y}}$ on a 2D rectangular lattice as an example, we can first consider the case where the inversion center is coincident with the glide line (Fig.~\ref{FigPT}(a)).  Because both $P$ and $\theta$ flip $\vec{k}$ and $P^{2}=+1$, all bands will be two-fold-degenerate for strong spin-orbit systems where $\theta^{2}=-1$.  However, to evaluate the potential semimetallic properties of this geometry, it is necessary to determine the $G_{y}$ eigenvalue structure of the bands along $\overline{\Gamma X}$ and $\overline{YM}$, as well as the locations of any required four-fold-degenerate points.  We can examine the relationship between $P,\theta$, and $G_{y}$ as representations on a four-site $\mathcal{H}(\vec{k})$ and use the additional factors of $e^{i\vec{k}\cdot\vec{t}}$ from the full lattice translations $\vec{t}$ to establish commutation relations.  Calling the representation of the operator for this case of inversion center location $G_{y}^{a}$, at a TRIM:

\begin{eqnarray}
G_{y}^{a}P &=& t_{x/2}M_{\hat{y}}P \nonumber \\
&=& Pt_{-x/2}M_{\hat{y}} \nonumber \\
&=& t_{x}Pt_{x/2}M_{\hat{y}} \nonumber \\
G_{y}^{a}P &=& e^{-ik_{x}}PG_{y}^{a}
\label{TRIMcom}
\end{eqnarray}

where in the final line we have acted $t_{-x}$ on an eigenstate of $G_{y}$.  For mathematical consistency, \emph{it was crucial that we have chosen all operations to have the same position-space origin} (here the inversion center).  Though this nuance can be overlooked when all mirror planes and lines are coincident with the inversion center, as they are in Ref.~\onlinecite{Steve2D},  \emph{it becomes a central detail when evaluating the eigenvalue character of systems for which one is unable to define a common origin for two spatial symmetries}, as we will see throughout this manuscript.  Equation~\ref{TRIMcom} implies that $\{G_{y}^{a},P\}=0$ at $X$ and $M$.  As $P^{2}=+1$ at those points, the emphasized statement in~\ref{sec:TRIMs} implies that representations at those TRIMs must be $4\times 4$, and thus that all states at $X$ and $M$ are four-fold-degenerate (Fig.~\ref{FigPT}(a)).  

Away from the TRIMs, we must additionally determine the $G_{y}$ eigenvalues of bands along  glide lines.  Considering $|+\rangle$ to be the positive eigenstate of $G_{y}$ such that:

\begin{equation}
G_{y}^{a}|+\rangle = ie^{ik_{x}/2}|+\rangle
\end{equation}

whose eigenvalue we compare to the local Kramers partner $P\theta|+\rangle$:

\begin{eqnarray}
G_{y}^{a}(P\theta|+\rangle) &=& t_{x/2}M_{\hat{y}}P\theta|+\rangle \nonumber \\
&=& P\theta t_{-x/2}M_{\hat{y}}|+\rangle \nonumber \\
&=& t_{x}P\theta G_{y}^{a}|+\rangle \nonumber \\
G_{y}^{a}(P\theta|+\rangle) &=& -ie^{ik_{x}/2}(P\theta|+\rangle) 
\end{eqnarray}

revealing that along both $\overline{\Gamma X}$ and $\overline{YM}$ all two-fold-degenerate bands have $G_{y}$ eigenvalues $\{+,-\}$ and thus can only anticross (Fig.~\ref{FigPT}(a)).  

However, as emphasized in Ref.~\onlinecite{ChenPT}, this picture changes significantly when the inversion center does not lie along a particular glide line or screw axis.  Consider a 2D rectangular system with a glide line $G_{y}=t_{x/2}M_{\hat{y}}$ that lies $a_{y}/4$ above the inversion center (Fig.~\ref{FigPT}(b)).  In order to consistently keep the commutation relations of the representations $[P,M_{i}]=0$ and $\{C_{2i},C_{2j}\}=-2\delta_{ij}$ (true for spinful systems where $(C_{2i})^{2}=-1$), we have to define at the operator level:

\begin{equation}
G_{y}^{b}=t_{y/2}t_{x/2}M_{\hat{y}} 
\end{equation}

where $G_{y}^{b}$ will here indicate a glide that lies a quarter-lattice $y$-direction displacement from the inversion center.  Reevaluating the commutation relations at the TRIMs:

\begin{eqnarray}
G_{y}^{b}P &=& t_{y/2}t_{x/2}M_{\hat{y}}P \nonumber \\
&=& Pt_{-x/2}t_{-y/2}M_{\hat{y}} \nonumber \\
&=& t_{x}t_{y}Pt_{x/2}t_{y/2}M_{\hat{y}} \nonumber \\
G_{y}^{b}P &=& e^{-ik_{x}}e^{ik_{y}}PG_{y}^{b}
\end{eqnarray}

where the final line is evaluated by acting the translations on an eigenstate of $G_{y}$.  In this case, $\{G_{y}^{b},P\}=0$ now at $X$ and $Y$, which by the arguments in~\ref{sec:TRIMs} requires that all states be four-fold-degenerate at those TRIMs (Fig.~\ref{FigPT}(b)).  

Moving off of the TRIMs, we can examine how this inversion-center offset affects the eigenvalue character of the local Kramers partners:

\begin{eqnarray}
G_{y}^{b}(P\theta|+\rangle) &=& t_{x/2}t_{y/2}M_{\hat{y}}P\theta|+\rangle \nonumber \\
&=& P\theta t_{-x/2}t_{-y/2}M_{\hat{y}}|+\rangle \nonumber \\
&=& t_{x}t_{y}P\theta G_{y}^{b}|+\rangle \nonumber \\
G_{y}^{b}(P\theta|+\rangle) &=& -ie^{ik_{x}/2}e^{ik_{y}}(P\theta|+\rangle). 
\end{eqnarray}

This implies that along $\overline{\Gamma X}$, two-fold-degenerate bands still have $G_{y}$ eigenvalue pairings $\{+,-\}$ and still always anticross.  But along $\overline{YM}$, if states are two-fold-degenerate, they will have $G_{y}$ eigenvalues $\{+,+\}$ or $\{-,-\}$, and along that glide line a four-fold crossing can therefore be locally protected (Fig.~\ref{FigPT})~\cite{ChenPT}.  However, \emph{we have not specified whether such a crossing must occur}, as is required for example in SrIrO$_{3}$ in space group 62~\cite{KeeRing}.  Such a distinction requires additional information about the nonsymmorphic symmetries present globally across the BZ.  In subsequent sections, we will provide both examples of layer group systems where four-fold crossings protected by an inversion-center offset are required in \emph{essential} semimetals and optional in \emph{band-inversion} semimetals.  

Two-fold-degenerate lines and planes can also occur in systems with two-fold nonsymmorphic symmetries~\cite{DDP,ChenNS}.  For example, in  a  rectangular two-dimensional system, the product of $\Pi=t_{x/2}M_{\hat{y}}$ and $\theta$ returns $\vec{k}$ to itself along $\overline{\Gamma Y}$ and $\overline{XM}$.  Independent of $\theta^{2}$, $(\Pi\theta)^{2}=t_{x}=e^{ikx}$ when acted on a state, guaranteeing states are at least two-fold-degenerate along $\overline{XM}$ where $(\Pi\theta)^{2}=-1$.  In these systems, bands along this line can be two-fold-degenerate and, as was the case with inversion symmetry, can be paired with either the same or the opposite eigenvalues of another symmetry operation valid along that line, depending on the relative origins of the crystalline symmetries.  This can, in at least the case of breaking a double Dirac point in space group 135, lead to an unusual Dirac semimetal \emph{without} inversion symmetry~\cite{DDP}.  In the 2D cases of the layer groups, there may not be enough degrees of freedom to create a system without inversion where such four-fold-degenerate crossings are the only features at the Fermi energy. 

Four-fold-degenerate lines are also occasionally possible in the layer groups.  For example, consider a system where along a line two crystalline symmetries $\Pi_{1}$ and $\Pi_{2}$ are valid and bands are already required to be two-fold-degenerate either by inversion and $\theta^{2}=-1$ or by a two-fold nonsymmorphic symmetry.  If $|+\rangle_{1}$ has the same $\Pi_{1}$ eigenvalue as its local Kramers partner and if $\{\Pi_{1},\Pi_{2}\}=0$ along this line, then that guarantees that $\Pi_{1}|+\rangle_{1}\neq\Pi_{1}\Pi_{2}|+\rangle_{1}$.  Therefore bands along the line are \emph{four-fold}-degenerate with $\Pi_{1}$ eigenvalues $\{+,+,-,-\}$.  

However, as exhaustively detailed in Ref.~\onlinecite{BigBook}, a system can only host one four-fold irreducible representation along a line in three dimensions, and therefore also in layer group systems.  Two four-fold-degenerate lines can thus never cross to form a locally-protected eight-fold-degenerate point along a line in three or fewer dimensions.    

\section{Semimetals in the Layer Groups}
\label{sec:LGs}

Seeking to examine nodal phenomena from both WPVZ bound and crystalline symmetry perspectives in strong spin-orbit systems in the layer groups, we present a simple model that typifies the layer group systems which contain essential or notable band-inversion nodal features.  In this section, we show how a four-site rectangular lattice can capture a large variety of both essential and band-inversion semimetallic features in two dimensions when its sites are dressed with three-dimensional vector objects, which one can think of as local displacements or dipole moments.  After presenting this construction, we present models for 7 specific layer groups, which represent all possible \emph{essential} semimetallic features in the layer groups, as well as relevant related examples of \emph{band-inversion} nodal features.  We sort our models by their WPVZ bounds in the platycosm formulation, and show using local eigenvalue character how nodal features are protected.  Finally, we show how some of these quasi-two-dimensional systems relate to existing three-dimensional semimetals.  

\begin{figure}
\centering
\includegraphics[width=3.5in]{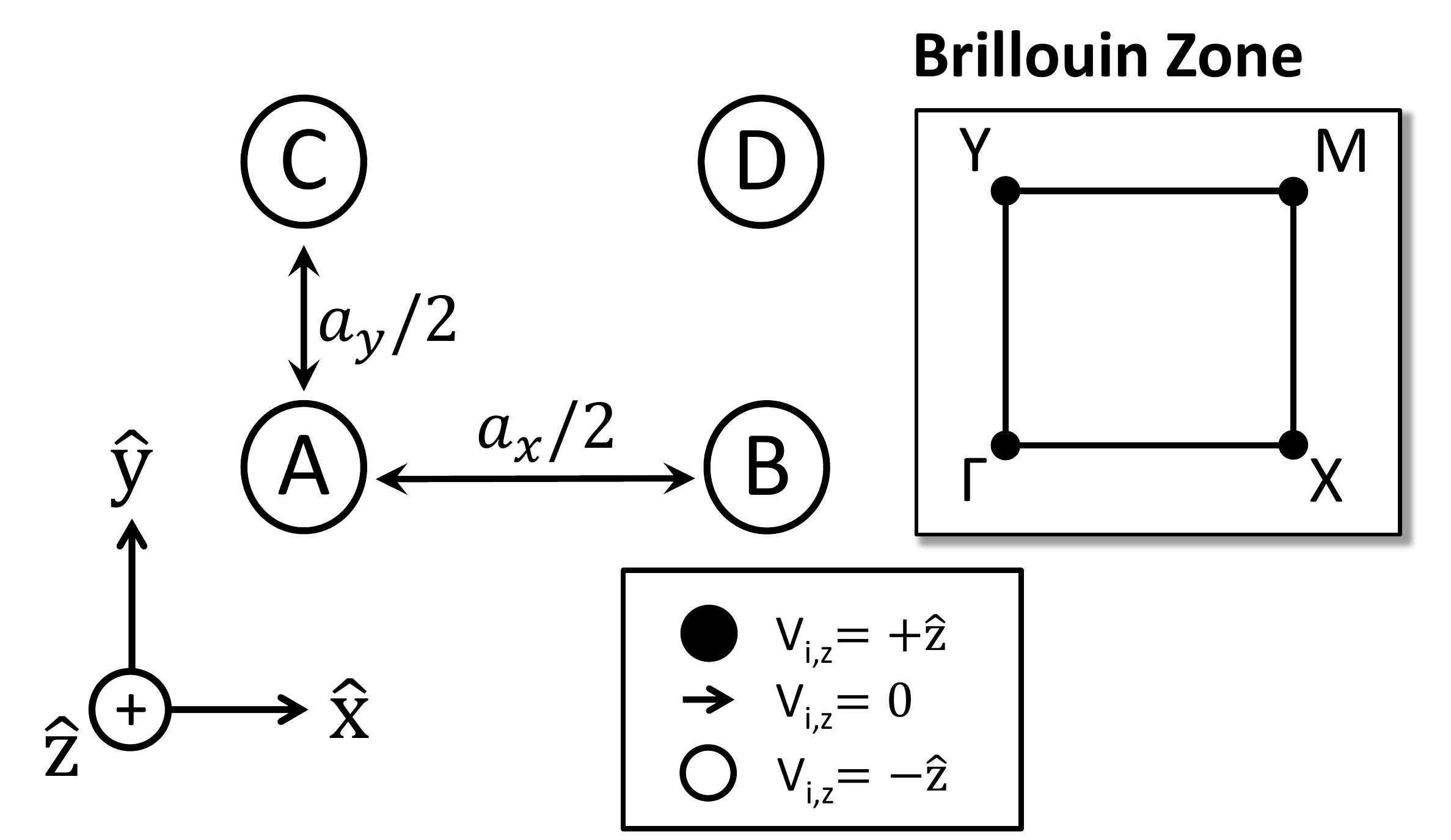}
\caption{Four-site model system for semimetals in the layer groups.  The four sublattices exist on a 2D rectangular unit cell in the $xy$ plane.  Each site has an additional spin degree of freedom.  Sites are then each dressed with a 3D vector object, visually represented by the symbols in the inset box, that relates them to their neighbors by a symmorphic symmetry operation and  a translation by half a lattice spacing $a_{x/y}/2$.  For real systems, this object can represent any time-reversal-symmetric property which transforms as a vector, such as displacement or a local dipole moment.  By selecting different A-site vectors and lattice generators, a diverse assortment of semimetallic phenomena can be realized.  Hamiltonians are generated by considering all first- and second-nearest neighbor hopping terms permitted by the restrictions imposed on $\mathcal{H}(\vec{k})$ by the generators of a particular layer group.  The specific terms allowed for each example layer group are detailed in Appendix~\ref{appendix:tb}.}
\label{FigLattice}
\end{figure}

As seen in Figure~\ref{FigLattice}, our model system consists of four sublattices arranged on a rectangular lattice.  Describing our sublattice space with Pauli matrices, $\tau^{x}$ describes s-orbital-like hopping between the A and B sites and $\mu^{x}$ describes s-orbital-like hopping between the A and C sites, such that second-neighbor s-like hopping is given by $\tau^{x}\mu^{x}$.  Each site also has a spin degree of freedom $\sigma$ which is flipped under time-reversal $\theta=i\sigma^{y}K\otimes(\vec{k}\rightarrow -\vec{k})$ for a general $\mathcal{H}(\vec{k})$.  This four-site unit cell, like that of any crystal system with multiple sites per unit cell, can be viewed as originating from a parent high-symmetry Bravais lattice, here a rectangular lattice with spherical sites.  Applying a time-reversal-symmetric tensor field (like an electric field) lowers the periodicity of the system and breaks some subset of point group symmetries at each site~\cite{MSGs}.  Our model can be realized by choosing just a dipole vector field, such that each site is dressed with a three-dimensional vector object that encodes the underlying crystal symmetries.  Physically, this vector can be considered as a local displacement of a single atom or a dipole moment between two atoms.

As one can observe by perusing the table in Appendix~\ref{appendix:fillingcond}, layer groups with $C_{3\hat{z}}$ or $C_{6\hat{z}}$ symmetries can only achieve WPVZ bounds of 2, owing to their inability to host nonsymmorphic symmetries.  While these systems can host rotation-protected band-inversion type semimetals, they will not be the focus of this paper.  One could consider a system with $C_{4\hat{z}}$ symmetry as a limiting case of the lattice in Figure~\ref{FigLattice}, in which case our model would collapse onto a version of the model in and recapture the physics of Ref.~\onlinecite{Steve2D}.  

Setting first all of the site vectors to $\vec{0}$, we can write down a very high-symmetry Hamiltonian consisting of all possible s-orbital-like hoppings between first- and- second nearest neighbors:

\begin{eqnarray}
\mathcal{H}_{0} &=& t_{x}\cos\left(\frac{k_{x}}{2}\right)\tau^{x} + t_{y}\cos\left(\frac{k_{y}}{2}\right)\mu^{x} \nonumber \\
&+&t_{2}\cos\left(\frac{k_{x}}{2}\right)\cos\left(\frac{k_{y}}{2}\right)\tau^{x}\mu^{x} 
\label{H_zero}
\end{eqnarray}

where we have set the lattice constants $a_{x}=a_{y}=1$ and enforce the inequivalence between $x$ and $y$ by keeping $t_{x}\neq t_{y}$.  As the on-site vectors are turned on, new hopping terms are allowed, and the symmetry is reduced into a particular layer group.  For a given layer group $LG$, the full second-neighbor Hamiltonian $H_{LG} = H_{0} + V_{LG}$, where $V_{LG}$ contains all of the layer-group-specific hopping terms beyond Eq.~\ref{H_zero}.  The details of deriving $V_{LG}$ for each of our examples, as well as layer-group-specific expressions for it, are noted in detail in Appendix~\ref{appendix:tb}.  

As detailed earlier in~\ref{sec:polygons}, by allowing placement onto at least one of the four platycosms in Figure~\ref{FigLayerPoly}, layer group systems can achieve WPVZ bounds of 2, 4, or 8, and will therefore host corresponding numbers of inseparably tangled bands.  Within the systems with WPVZ bounds of 2 and 4, band-inversion metallic and nodal features are also possible for this eight-band model.  In the following sections, we present typifying examples for layer group semimetals within each possible WPVZ bound, showing for each example how the WPVZ bound relates to the more familiar crystalline symmetry analysis.   

\subsection{WPVZ Bound of 2}

Without the presence of a nonsymmorphic symmetry, only time-reversal symmetry can force bands to group together~\cite{WPVZ,SaadDirac,Steve2D}.  Layer groups with only symmorphic symmetries have band structures with two-fold degeneracies at the time-reversal-invariant momenta when $\theta^{2}=-1$.  Therefore, at even fillings, any such system with more than two bands can either be an insulator or a band-inversion semimetal.  In the layer groups, there are myriad ways for a band-inversion crossing to be locally protected by mirror or rotation eigenvalues, so we will only explore one such example in a system with singly-degenerate bands.  

To start, consider a simple system for which the $xy$-plane itself is a mirror, such that all bands have good $M_{\hat{z}}$ quantum numbers.  For real materials this corresponds to a system which is not buckled or has no additional stacking or external field structure which distinguishes $\pm\hat{z}$, such as graphene without a substrate.  

\begin{figure}
\centering
\includegraphics[width=3.5in]{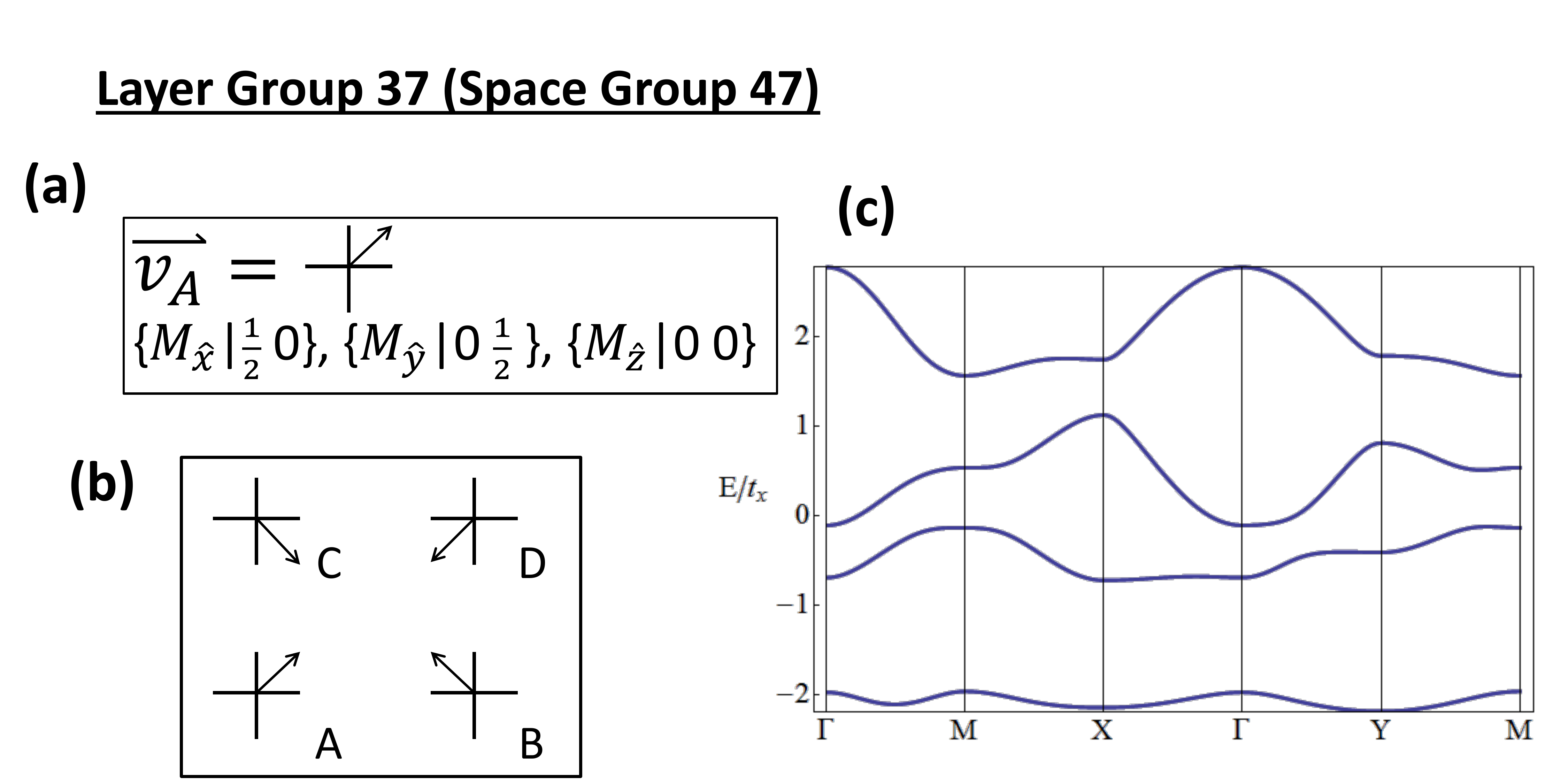}
\caption{The generators (a), lattice (b), and a typical band structure (c) for \emph{pmmm}, layer group 37, (space group 47).  All elements of the layer group are symmorphic, with mirror lines separating the A and B sublattices and the A and C sublattices.  As the $xy$-plane itself is a mirror, this system is flat and has inversion symmetry, with the inversion center lying at the center of the four sites and at the intersection of all 3 mirror lines and planes.  Therefore, by the arguments in~\ref{sec:multiplets}, bands are two-fold-degenerate with $M_{\hat{z}}$ eigenvalues $\{+,-\}$.  Consequently, the bands can only anticross, and at even fillings this system is always an insulator (c).}
\label{FigLG37}
\end{figure}

Restricting ourselves to one of these flat layer groups, layer group 37 \emph{pmmm} (which when stacked, is equivalent to space group 47), we first choose an example with only symmorphic symmetries, and therefore a WPVZ bound of 2.  The presence of inversion symmetry P, combined with time-reversal makes \emph{all} bands two-fold-degenerate (Fig.~\ref{FigLG37}). The inversion center lies at the intersection of all three mirror lines and planes (Fig.~\ref{FigLG37}(b)), and therefore all states and their local Kramers partners have opposite $M_{\hat{z}}$ eigenvalues, as detailed in~\ref{sec:multiplets} and Ref.~\onlinecite{ChenPT}.  Therefore, regardless of band-tuning conditions, this system will generically be an insulator, because nearby bands can only anticross (Fig.~\ref{FigLG37}(c)).  

\begin{figure}
\centering
\includegraphics[width=3.5in]{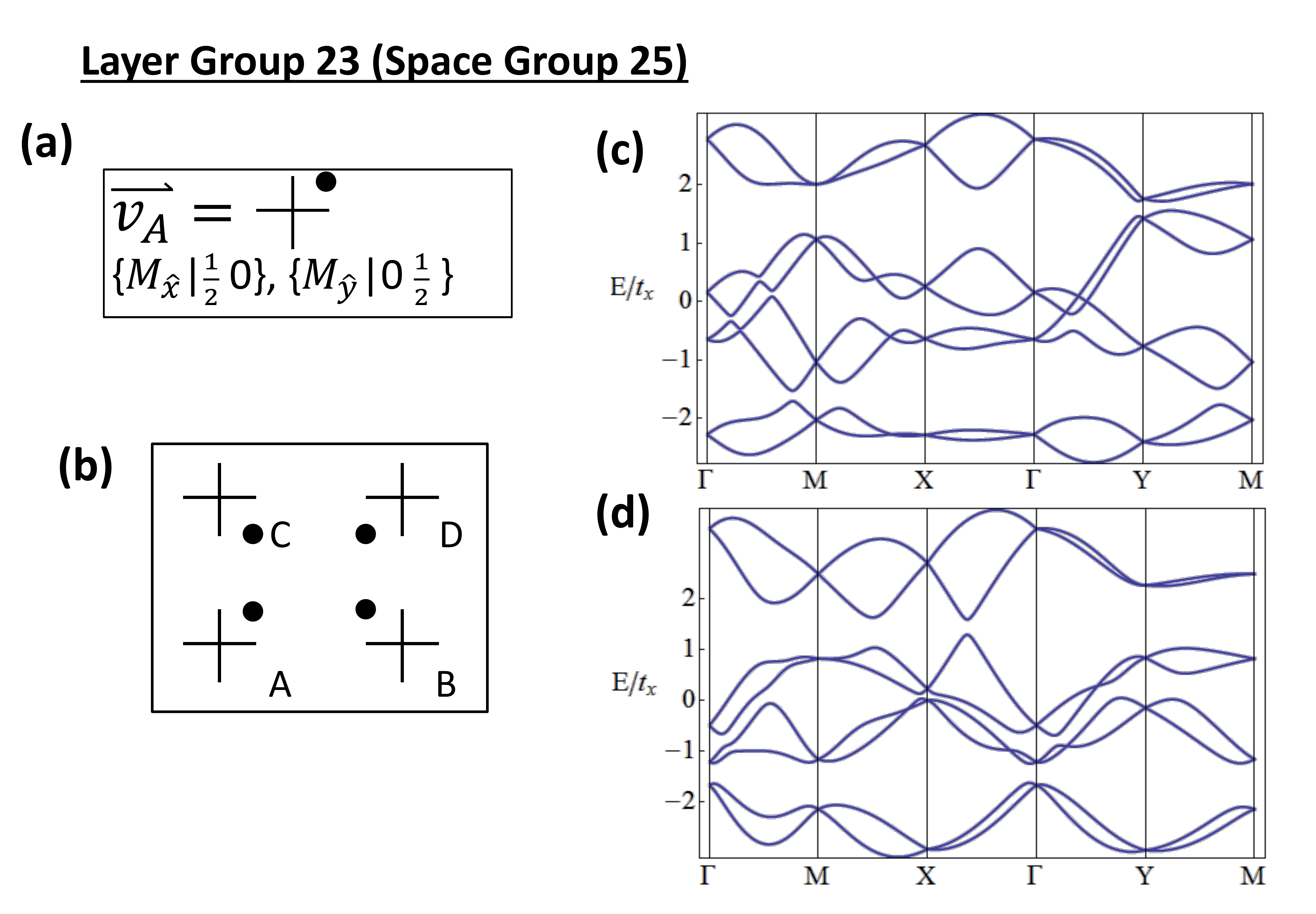}
\caption{The generators (a), lattice (b), and a typical band structure (c) for \emph{pmm2}, layer group 23, (space group 25).  All elements of the layer group are symmorphic, with mirror lines separating the A and B sublattices and A and C sublattices.  The vectors have been bent up into the $+\hat{z}$ direction, breaking $M_{\hat{z}}$ as one would see if there were a substrate or a perpendicular electric field added to a system in layer group 37 (Fig.~\ref{FigLG37}).  Consequently, this system is also a \emph{wallpaper group}, and could describe the surface of a three-dimensional object.  Without inversion, nonsymmorphic symmetries, or $n>2$ $C_{n\hat{z}}$ rotation points, bands are singly degenerate and can only cross with local protection by mirror eigenvalues on the mirror lines.  Typical values of the tight-binding parameters give metallic states at half-filling (c), but values can also be chosen to separate the bands into groups of two and open up consistent gaps at all even fillings (d).}  
\label{FigLG23}
\end{figure}

Conversely, one could imagine putting on an electric field, like that of a substrate, which bends the on-site vectors out of the plane in the $+\hat{z}$ direction.  As shown in Figure~\ref{FigLG23}, this reduces the layer group to 23 \emph{pmm2} (space group 25) and breaks $M_{\hat{z}}$ and $P$, allowing for new first- and second-nearest-neighbor hopping terms.  As this layer group only consists of in-plane mirrors and rotations about the $z$ axis, it is also one of the \emph{wallpaper} groups described in~\ref{sec:polygons} and could be constructed in purely two dimensions, for instance as the surface of a three-dimensional object.  Bands in this layer group are now singly degenerate, and therefore have the ability to cross with local protection.  Consider tuning the tight binding parameters for this layer group to roughly physical values, such that first-neighbor hopping terms are larger than second-neighbor ones and s-orbital-like hopping terms are larger than the terms for spin-orbit interaction.  For these typical values of the tight-binding parameters, this system is a band-inversion semimetal (Fig.~\ref{FigLG23}(c)).  However, values could also be chosen to open up gaps at all even fillings (Fig.~\ref{FigLG23}(d)), because in this layer group, the WPVZ bound only requires that bands tangle together in groups of 2. 

\subsection{WPVZ Bound of 4}

\begin{figure}
\centering
\includegraphics[width=3.5in]{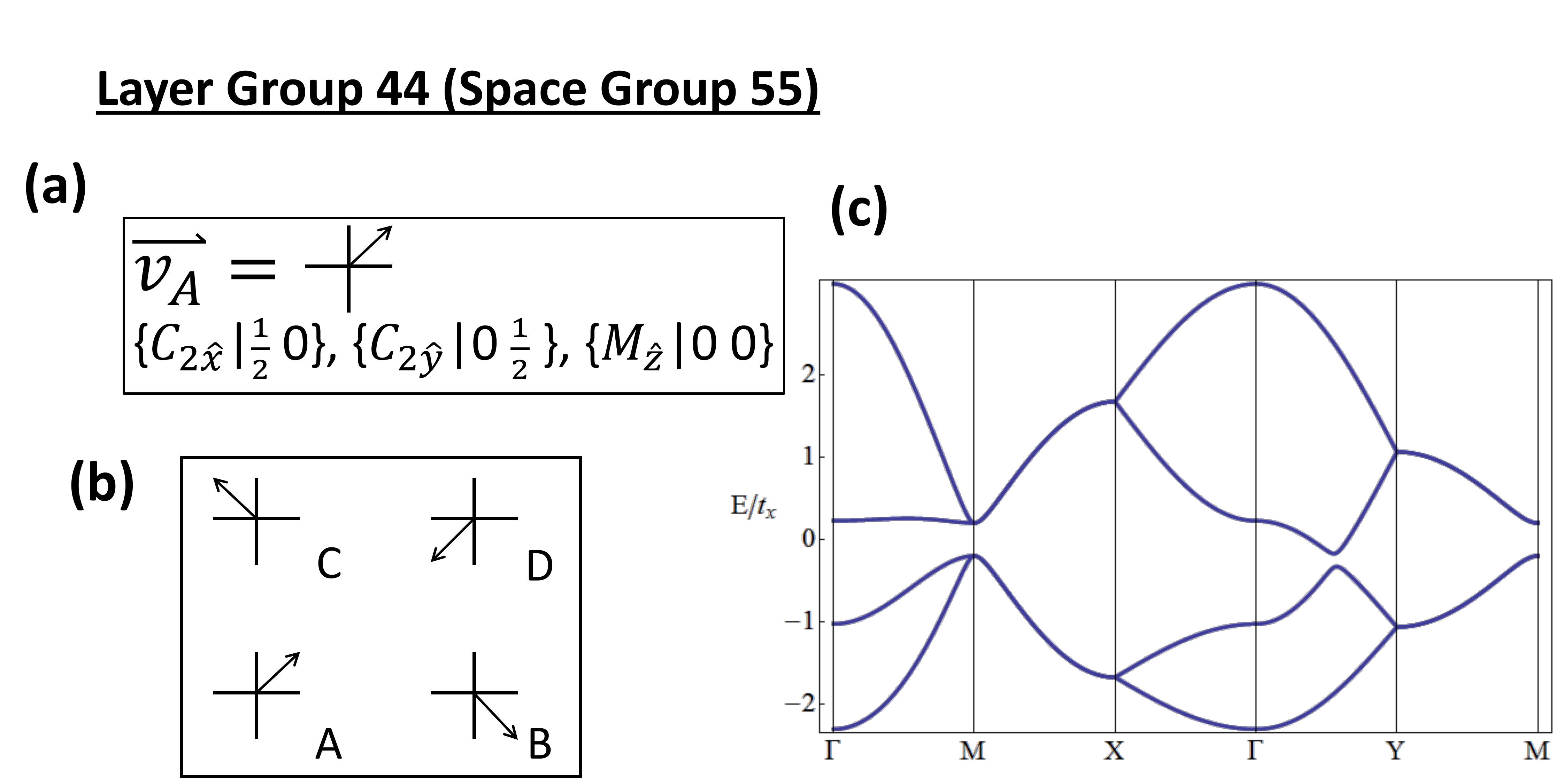}
\caption{The generators (a), lattice (b), and a typical band structure (c) for \emph{pbam}, layer group 44, (space group 55).  This group is a flat layer group generated by constraining  into the $xy$-plane a system generated by two perpendicular screws.  Consequently, it has inversion symmetry, with the inversion center located in the center of the unit cell, off of the glide lines resulting from the product of $P$ and $S_{2x/y}=t_{x/y/2}C_{2x/y}$.  All bands are at least two-fold-degenerate by $(P\theta)^{2}=-1$ and bands along $\overline{XM}$ and $\overline{YM}$ are four-fold-degenerate by the combination of glide mirror and $M_{\hat{z}}$, as detailed in~\ref{sec:multiplets}.  Bands at $X$, $Y$, and $M$ are four-fold-degenerate due to the relationship between $P$ and $S_{2x/y}$, as detailed in~\ref{sec:TRIMs} and Ref.~\onlinecite{Steve2D}, and disperse linearly.  Therefore, at fillings of $\nu=2,6$, this system is an essential Dirac line node semimetal.  Two-fold-degenerate bands are all paired with $M_{\hat{z}}$ eigenvalues $\{+,-\}$ or in four-fold multiplets with $M_{\hat{z}}$ eigenvalues $\{+,+,-,-\}$, and therefore cannot be tuned to cross by band inversion.  Consequently, at half filling ($\nu=4$), this system is necessarily an insulator.}  
\label{FigLG44}
\end{figure}

A layer group system with one or more two-fold nonsymmorphic symmetries is allowed placement onto a platycosm other than the 3-torus, and it will therefore host essential groupings of four or eight bands~\cite{WPVZ}.  Within systems with a WPVZ bound of 4, band-inversion semimetals are still possible between groupings of 4 bands, though if bands are two-fold-degenerate, additional conditions are required for determining if band inversions can be locally protected by eigenvalue character.  In this section, we first examine two layer groups with WPVZ bounds of 4, which at fillings of $\nu=2,6$ are essential semimetals with Weyl or Dirac features, as explored in Ref.~\onlinecite{Steve2D}.  After, we present an example of a band-inversion Dirac semimetal at half filling, protected locally by the inversion-center-offset arguments in~\ref{sec:multiplets} and Ref.~\onlinecite{ChenPT}.  

We start with a high-symmetry flat system in layer group 44 \emph{pbam} (space group 55) (Fig.~\ref{FigLG44}).  This group is generated by two perpendicular screws protruding from the sites enforced in combination with $M_{\hat{z}}$, such that the site vectors are confined in the $xy$ plane (Fig.~\ref{FigLG44}(a)).  For determining the WPVZ bound, one could either choose a screw and mod out onto the dicosm, or combine a screw with $M_{\hat{z}}$ to create a glide line and mod out onto the 1st amphicosm:

\begin{equation}
t_{x/2}C_{2x}M_{\hat{z}} = t_{x/2}C_{2x}PC_{2z} = -it_{x/2}PC_{2y}=-i(t_{x/2}M_{\hat{y}})
\end{equation}

where we have used the fact that in spinful systems the representations of two-fold rotations obey the same algebra as the Pauli matrices.  Either choice of manifold results in a WPVZ bound of 4, a property clearly visible by noting the gap at $\nu=4$ in the band structure (Fig.~\ref{FigLG44}(c)).  

Due to the relationship between two-fold nonsymmorphic symmetries and inversion, highlighted in~\ref{sec:TRIMs} and Ref.~\onlinecite{Steve2D}, the TRIMs at the end of the translation directions (X,Y, and M) all host four-fold degeneracies.  All bands are at least two-fold-degenerate with pairs of opposite $M_{\hat{z}}$ eigenvalues.  Bands along $\overline{XM}$ and $\overline{YM}$ have the requisite offset from the inversion center to have screw eigenvalues $\{+,+\}$, but due to the additionally valid $M_{\hat{z}}$ are four-fold-degenerate with mirror eigenvalues $\{+,+,-,-\}$, as detailed in~\ref{sec:multiplets}, and are thus unable to cross.  Therefore, at half filling ($\nu=4$), this system is always an insulator.  Observing symmetry-allowed terms in the tight-binding model in Appendix~\ref{appendix:tb}, all four-fold points and lines disperse linearly, and therefore at fillings of $\nu=2,6$ this system is an essential Dirac line node semimetal.  

\begin{figure}
\centering
\includegraphics[width=3.5in]{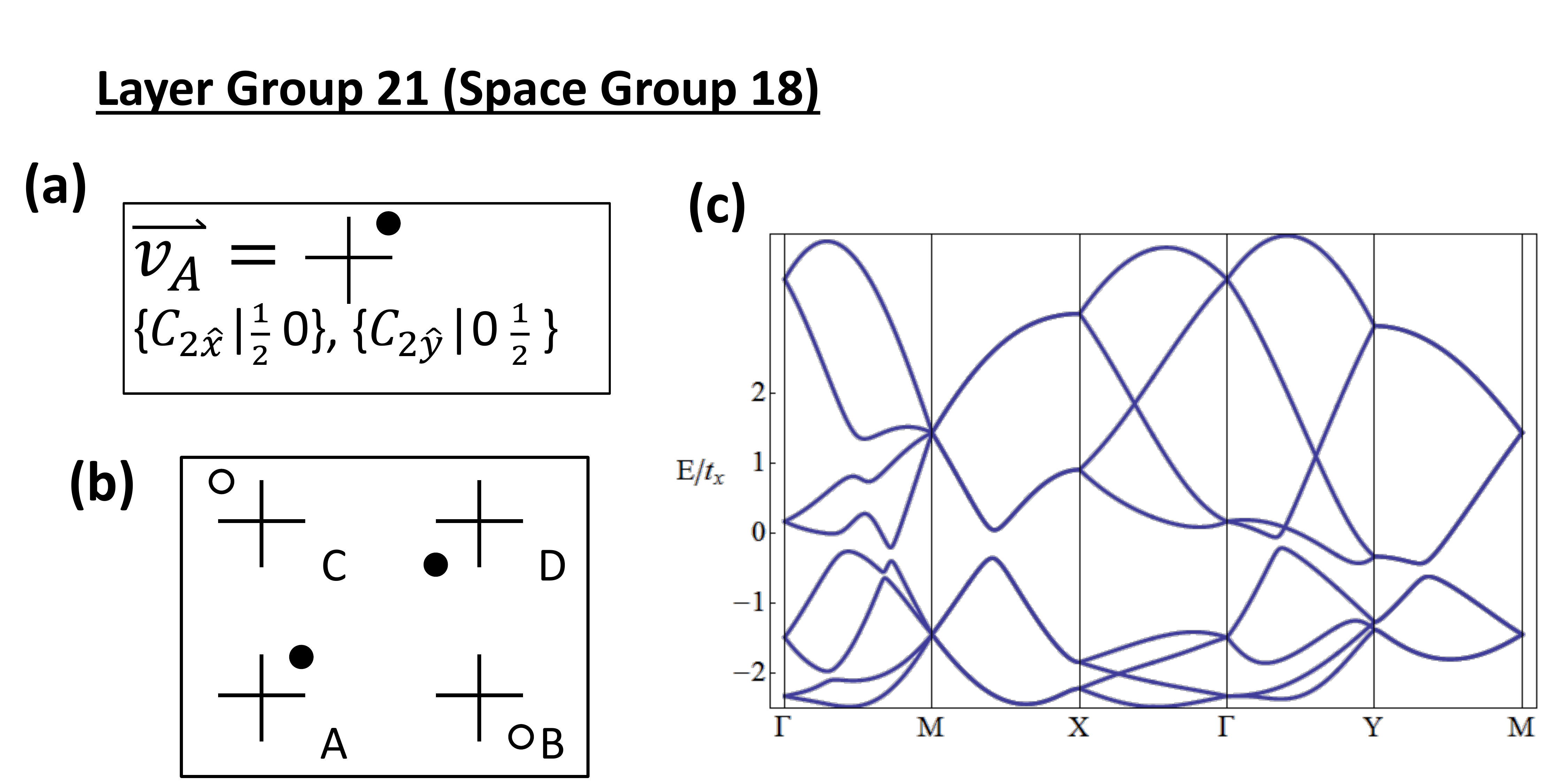}
\caption{The generators (a), lattice (b), and a typical band structure (c) for \emph{p$\textit2_{1}\textit2_{1}$2}, layer group 21, (space group 18).  This low-symmetry group is generated just by two perpendicular two-fold screws protruding from the sites.  Due to having broken inversion symmetry and a combination of nonsymmorphic symmetries incompatible with placement onto the 1st amphidicosm in Fig.~\ref{FigLayerPoly}, this system has essential 4-band tangles which resemble hourglasses, such that at fillings of $\nu=2,6$, it has 2D  essential Weyl points along $\overline{\Gamma Y}$ and $\overline{\Gamma X}$~\cite{Steve2D}.  Bands along most lines are singly degenerate and therefore capable of crossing with symmetry protection by the same mechanism as in Fig.~\ref{FigLG37} (though for the choice of parameters in (c) the system is an insulator at half filling; $\overline{\Gamma Y}$ is narrowly gapped).  Bands along $\overline{XM}$ and $\overline{YM}$ are two-fold-degenerate by the combination of a two-fold nonsymmorphic symmetry and $\theta$, as detailed in~\ref{sec:multiplets}.  M hosts four-fold points despite the absence of inversion, owing to $\{S_{2x},S_{2y}\}=0$ and $(S_{2x})^{2}=(S_{2y})^{2}=+1$ at this point, as detailed in~\ref{sec:TRIMs}.  At fillings of $\nu=2,6$, this system is therefore an essential semimetal, and can be tuned to have a minimal Fermi surface of four Weyl points and a Dirac point.}  
\label{FigLG21}
\end{figure}

We can locally break $M_{\hat{z}}$ on the A site and then use the same screws as generators to produce layer group 21 \emph{p$\textit2_{1}\textit2_{1}$2} (space group 18) (Figure~\ref{FigLG21}).  In this system, bands are now singly degenerate, except along $\overline{XM}$ and $\overline{YM}$ where they are paired by the combination of $\theta$ and a two-fold nonsymmorphic symmetry, as detailed in~\ref{sec:multiplets}.  As verified by analysis of symmetry-allowed linear terms in the tight binding model in Appendix~\ref{appendix:tb}, bands along $\overline{\Gamma X}$ and $\overline{\Gamma Y}$ feature essential 2D Weyl points and form essential 4-band tangles which resemble hourglasses~\cite{Steve2D}.  As noted in~\ref{sec:nonsym}, these points cannot be paired at any of the TRIMs and eliminated as long as the two screws are preserved, as the combination of the nonsymmorphic symmetry and $\theta^{2}=-1$ provides a topological obstruction to doing so, even when this system is stacked into the third dimension and the Weyl points become 3D with well-defined Chern numbers.  At M, the two screws anticommute and square to $+1$, and therefore despite the absence of inversion symmetry, bands at M are four-fold-degenerate, and in fact are linearly dispersing.  Therefore, at fillings $\nu=2,6$, this system is an essential point node semimetal, and can be tuned to have a minimal Fermi surface of four Weyl points and a Dirac point.  Though we have chosen parameters which gap this system at half filling (Fig.~\ref{FigLG21}(c)), the singly-degenerate bands across many of the high-symmetry lines are capable of inverting with local protection by the same mechanism as a previous system with a WPVZ bound of 2, layer group 23 (Fig.~\ref{FigLG23}).  

It is worth noting that despite having two perpendicular nonsymmorphic symmetries, layer groups 44 and 21 do not achieve WPVZ bounds of 8.  One can understand this bound limitation by recalling that a key requirement of the modding procedure in~\ref{sec:polygons} was the selection of manifolds without fixed points.  In terms of symmetry operations, this requirement can be restated, as noted by Bieberbach, that \emph{the product of the nonsymmorphic symmetries selected for modding, as defined from a common origin, must itself also be a nonsymmorphic operation}~\cite{Conway}.  Examining the two screw generators:

\begin{equation}
t_{x/2}C_{2x}t_{y/2}C_{2y}=it_{x/2}t_{-y/2}(C_{2z})
\end{equation}

which is itself just a $C_{2z}$ symmorphic rotation about the center of the unit cell.  As we will see in the subsequent section, the only combination of operations which can achieve a WPVZ bound of 8 in layer group systems is $t_{x/2}M_{\hat{z}}$ and $t_{y/2}C_{2y}$.   

\begin{figure}
\centering
\includegraphics[width=3.5in]{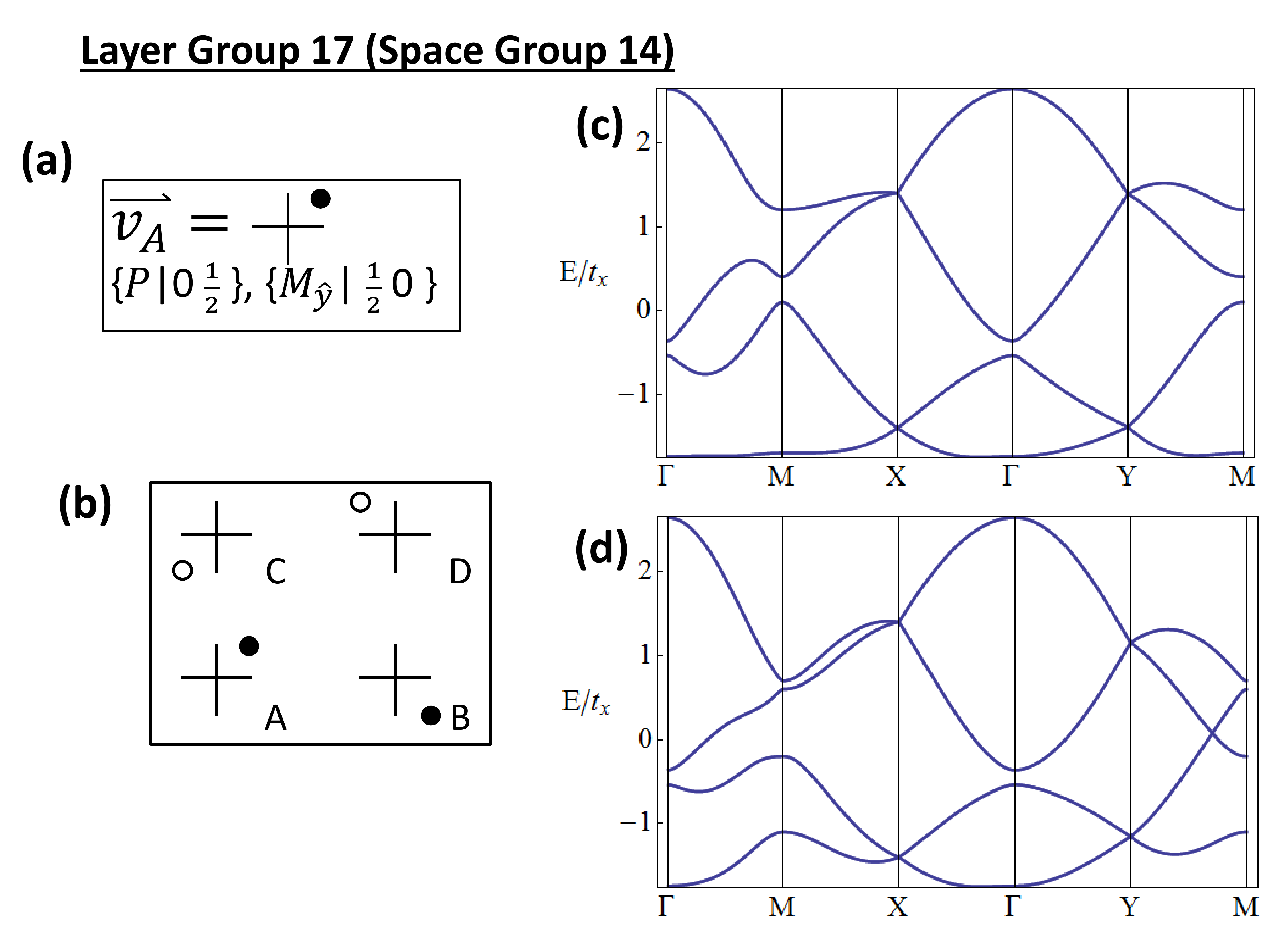}
\caption{The generators (a), lattice (b), and two possible band structures (c,d) for \emph{p$\textit2_{1}/b\textit1\textit1$}, layer group 17, (space group 14).  The lattice has horizontal glide lines along the sites and inversion centers between the A and C sites and between the B and D sites.  Due to the combination of P and $\theta$, bands everywhere are two-fold-degenerate.  The offset between the inversion centers and the glide lines leads to four-fold degeneracies at $X$ and $Y$, as noted in Fig.~\ref{FigPT} and in~\ref{sec:multiplets}.  As these four-fold points are linearly dispersing, in accordance with the WPVZ bound this system is an essential Dirac semimetal at fillings of $\nu=2,6$, allowing an idealized Fermi surface consisting of two Dirac points.  At half filling, however, this system is capable of being both an insulator (c) or a semimetal (d), as bands along $\overline{YM}$ are two-fold-degenerate with pairs of the same glide mirror eigenvalue.  This semimetallic phase is locally protected by the statements in Ref.~\onlinecite{ChenPT}, but can be gapped out by a band-inversion transition.  Therefore, for guaranteeing the existence of an \emph{essential} 8-band Dirac semimetallic phase, like that in SrIrO$_3$ in Ref.~\onlinecite{KeeRing}, we find the inversion-center offset highlighted in Ref.~\onlinecite{ChenPT} to be a necessary, but insufficient condition, and that we must require additional constraints.}  
\label{FigLG17}
\end{figure}

Within the layer groups with WPVZ bounds of 4, one can also achieve a Dirac semimetal at fillings $\nu\in4\mathbb{Z}$ through a band inversion transition.  In \emph{p$\textit2_{1}/b\textit1\textit1$}, layer group 17, (space group 14) (Fig.~\ref{FigLG17}), there exists an offset between horizontal glide lines, which connect adjacent sites A and B, and inversion centers, which lie between sites A and C (Fig.~\ref{FigLG17}(b)).  As explained in Ref.~\onlinecite{ChenPT} and in~\ref{sec:multiplets}, this offset allows bands along $\overline{YM}$ to be two-fold-degenerate with pairs of the same glide mirror eigenvalue.  At fillings $\nu=2,6$, the tight-binding models in Appendix~\ref{appendix:tb} show that this system is an essential Dirac semimetal with Dirac points at $X$ and $Y$, but also show that other nodal features are possible at $\nu=4$.  A band inversion about a TRIM, here $M$, leads to the creation of a Dirac point along $\overline{YM}$ and its time-reverse (Fig.~\ref{FigLG17}(d)).  

In a three-dimensional stack of this system, this feature would instead emerge as a Dirac line node.  However, unlike the Dirac line node in SrIrO$_3$ in Ref.~\onlinecite{KeeRing}, which is also locally protected by an inversion-center offset and a glide mirror, this line node in space group 14  could be removed by a band-inversion transition.  As we will examine in the next section, SrIrO$_3$ in space group 62 has a WPVZ bound of 8 and is in fact more closely related to a different layer group.  Therefore, we find that for the protection of essential 8-band Dirac nodal features such as the line node in SrIrO$_3$, the statements from Ref.~\onlinecite{ChenPT} \emph{are necessary for local protection, but insufficient for guaranteeing existence.}  As we will explore in the following section, an inversion-center offset from a nonsymmorphic symmetry is only one of three conditions required to form an essential Dirac line node.  In fact, we will see that 8-band essential semimetallic structures can even be formed \emph{in the absence of inversion symmetry.}  

\subsection{WPVZ Bound of 8}
\label{WPVZ8}

\begin{figure}
\centering
\includegraphics[width=3.5in]{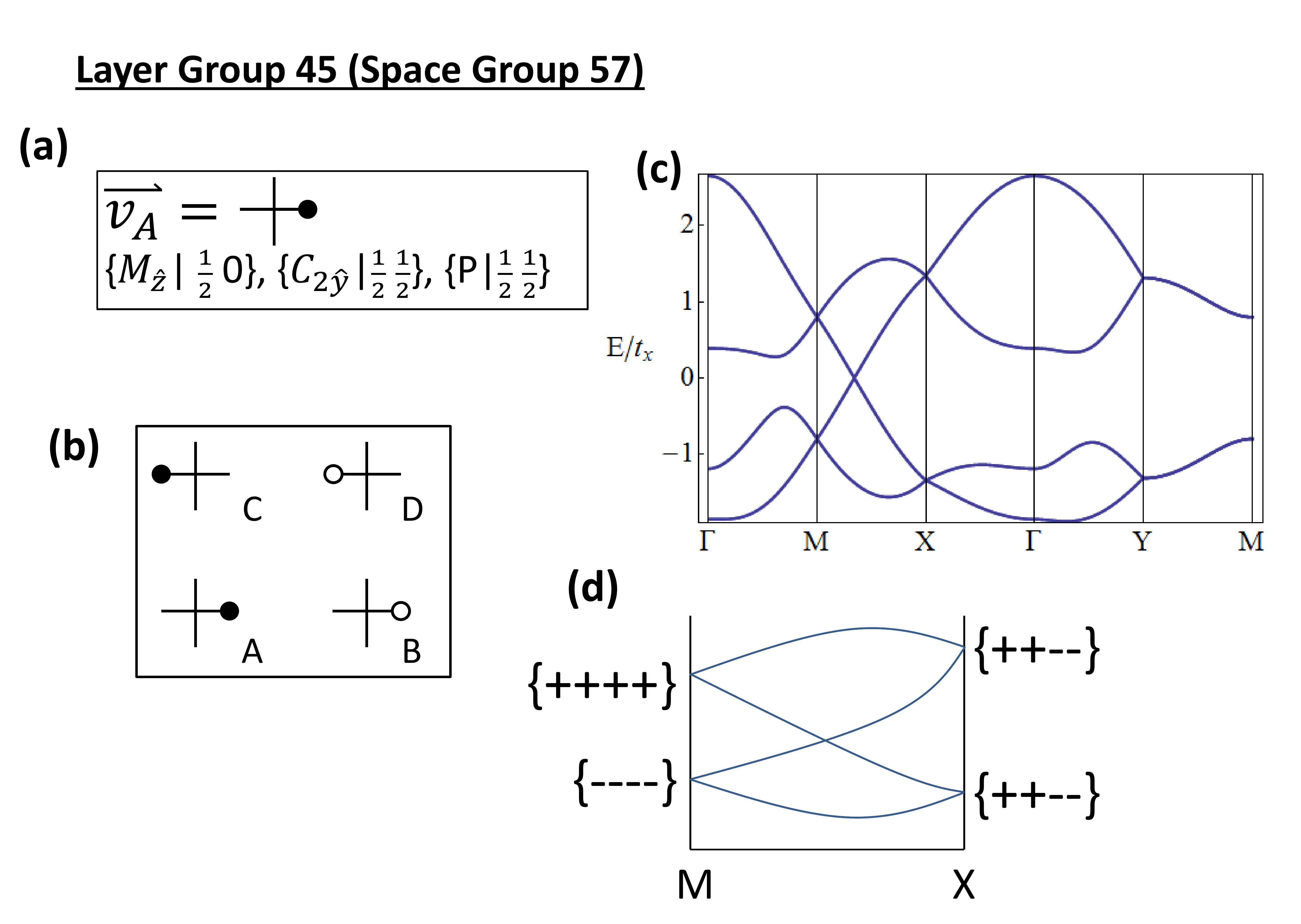}
\caption{The generators (a), lattice (b), and a typical band structure (c) for \emph{pbma} layer group 45 (space group 57).  This high-symmetry layer group has glide mirrors in the $x$ and $z$ directions and $M_{\hat{y}}$ about the sites, such that it has an inversion center in the center of its unit cells, off of glide line $G_{x}=t_{y/2}M_{\hat{x}}$.  This offset allows for bands along $\overline{XM}$ to be two-fold-degenerate with the same $G_{x}$ eigenvalues, which locally protects Dirac points along that line and its time-reverse, much like the local protection of the Dirac point in Fig.~\ref{FigLG17}(d).  However, unlike in that previous semimetal, whose nodal features were optionally created by tuning through a band-inversion transition, the Dirac points in layer group 45 are \emph{essential}, making them more like the essential Dirac line node in SrIrO$_3$ in space group 62~\cite{KeeRing}.  Because four-fold points are required at $X$ and $M$ by the relationship between $G_{x}$, $S_{y}=t_{x/2}t_{y/2}C_{2y}$, and $P$, and because $G_{x}$ commutes with all other independent symmetry operations of the layer group at $M$, four-fold points at $X$ and $M$ have differing $G_{x}$ eigenvalue pairings (d), leading to the required crossing along $\overline{MX}$.}  
\label{FigLG45}
\end{figure}

Though many layer groups exist with multiple perpendicular, two-fold nonsymmorphic symmetries, only 3 such groups exist which satisfy the condition for modding out more than one nonsymmorphic operation, namely that \emph{the product of the two operations, as defined from a common origin, is itself also a nonsymmorphic operation}~\cite{Conway}.  We find that in the 80 layer groups, only groups with both $t_{x/2}M_{\hat{z}}$ and $t_{y/2}C_{2y}$ (as well as any trivial, in-plane rotations of them) can achieve this condition and therefore allow placement onto the 1st amphidicosm and have WPVZ bounds of 8.  

Of these three layer groups, two of them, layer groups 43 and 45, have inversion symmetry and are very similar to each other.  Choosing to focus on the high-symmetry layer group 45, \emph{pbma} (Fig.~\ref{FigLG45}), we can see clearly along $\overline{XM}$ a robust 8-band Dirac feature which matches the nodal ring in SrIrO$_3$~\cite{KeeRing} (Fig.~\ref{FigLG45}(c)).  In fact, space group 57, the stacked equivalent of layer group 45, is closely related to the SrIrO$_3$ $pbnm$ space group 62, with the chief difference coming from the substitution of a three-dimensional ``n-glide'' with an in-plane ``b-glide''.   

The relationship between layer group 45 and space group 62 can be examined both from a consideration of allowed flat manifold placements and from an evaluation of band multiplicity and symmetry eigenvalue structure.  In the language of WPVZ, layer group 45 (space group 57) has a four-site unit cell with nonsymmorphic symmetries $t_{x/2}M_{\hat{z}}$ and $t_{y/2}C_{2y}$ as defined from the common origin of the midpoint between the A and B sites, leading as explained in~\ref{sec:polygons} to an allowed placement onto the 1st amphidicosm and the requirement that at least 8 bands be tangled together.  Space group 62 has a four-site unit cell with its $S_{y}$ above the glide plane, so using the same axes one can mod out the nonsymmorphic operations $t_{x/2}M_{\hat{z}}$ and $t_{z/2}t_{y/2}C_{2y}$, allowing placement onto the 2nd amphidicosm and requiring that at least 8 bands be tangled together (the 2nd amphidicosm is a fundamentally three-dimensional manifold and does not neatly decompose into the modified fundamental polygons from Fig.~\ref{FigLayerPoly}.)  Finally, as neither space group 57 nor space group 62 is among the 10 known space groups for which the platycosm formulation of the WPVZ bound is insufficient, we can deduce that in these two space groups all minimally tangled bands will in fact come in groups of exactly 8~\cite{WPVZ}.  

From a symmetry perspective, the story regarding this band structure is a bit more involved.  We find that there are three criteria which must all be met to guarantee the existence of an essential eight-band tangle in an orthorhombic system with inversion symmetry.  First, one must check whether or not any two-fold nonsymmorphic symmetry lines or planes exist offset from an inversion center.  As detailed in~\ref{sec:multiplets} and Ref.~\onlinecite{ChenPT}, a system which fulfills this criterion, if it has two-fold-degenerate bands along the $k$-space nonsymmorphic line or plane at the zone boundary, will have multiplets with the same eigenvalue of the nonsymmorphic symmetry that can cross and form Dirac points with local protection.  The second criterion involves checking whether such a crossing is prohibited from being removed by a band-inversion transition.  Such a process can only be prevented when bands at the TRIMs on either side of the two-fold nonsymmorphic line or plane, here at $M$ and at $X$, are all four-fold-degenerate.  Determining this degeneracy can either be accomplished by considering at those TRIMs the algebra between \emph{all} of the independent space group generators, or by consulting a crystalline symmetry textbook such as Bradley and Cracknell~\cite{BigBook}.  Finally, and most importantly, to prove that an odd number of Dirac crossings \emph{must} exist along the high-symmetry line, one must additionally show that the four-fold-degenerate bands at the bounding TRIMs support a particular algebra.  Specifically, at the TRIM where the nonsymmorphic operation which is offset from the inversion center squares to $+1$, \emph{that symmetry must commute with all other independent operations which generate the space group and are valid at that TRIM.}  If this final criterion is met, then the nonsymmorphic operation must have at that point, here $M$, a $4\times 4$ representation proportional to the identity and therefore eigenvalues $\{+,+,+,+\}$ or $\{-,-,-,-\}$.  As at the other bounding TRIM where the nonsymmorphic operation squares to $-1$, here $X$, time-reversal requires that the imaginary nonsymmorphic eigenvalues be paired $\{+,+,-,-\}$, doubly-degenerate bands must cross an odd number of times between the two bounding TRIMs and form at least one \emph{essential} Dirac crossing.  In three dimensions, if this occurs in a glide plane, then any path between the two TRIMs must contain a Dirac crossing and a Dirac line node forms, as is the case in SrIrO$_3$.  Elements of these criteria for this particular iridate system were recently noted in Ref.~\onlinecite{KeeExplain}.  

Returning to layer group 45, we can examine how evaluating these criteria works in practice.  Only the product of two generators, $G_{x}=t_{y/2}M_{\hat{x}}$, satisfies the inversion-center-offset criterion and is therefore suitable for consideration.  Bands at $X$ must be four-fold-degenerate, as $\{G_{x},P\}=0$ here and $P^{2}=+1$, as noted in~\ref{sec:TRIMs}.  At $M$, another two-fold operation, such as $t_{y/2}C_{2y}$, as defined from an inversion center, can anticommute with one of the two remaining independent layer group generators and mandate that bands there also be four-fold-degenerate.  Finally, we can note that as defined from a common origin along the $G_{x}$ lines, $[G_{x},\Pi]=0$ at $M$ for all $\Pi$, where $\Pi$ is an independent generating operation of layer group 45. Matching bands with the same $G_{x}$ eigenvalues $\lambda^{\pm}=\pm ie^{ik_{y}/2}$, Fig.~\ref{FigLG45}(d) shows that these criteria necessitate the existence of a Dirac point along $\overline{XM}$.   

While more involved than the WPVZ bound method for determining if groups of eight bands have to be tangled together in a system with inversion symmetry, this consideration of symmetry eigenvalues and commutation relations is beneficial when dealing with three-dimensional space groups for which the platycosm formulation of the WPVZ bound breaks down.  For example, in systems where high-fold rotation leads to an eight-fold double Dirac point, the WPVZ bound calculated through a consideration of flat manifold placement only seems to capture the particular 8-band Dirac feature seen in layer group 45.  While the platycosm formulation of the WPVZ bound states that space group 130 has a minimal insulating filling of $8\mathbb{Z}$ and space group 135 a minimal insulating filling of $4\mathbb{Z}$, both crystal systems are very similar in practice and both in practice host essential double Dirac points~\cite{DDP}.  The only distinguishing feature between them is that additional Dirac points are present in space group 130 at fillings $\nu\in4+8\mathbb{Z}$, owing to the offset of a screw rotation from the inversion centers.  The same set of two-fold nonsymmorphic symmetries lies \emph{along} the inversion centers in space group 135, and therefore the first part of the criteria for local protection of such Dirac points is not met.  

This consideration of two-fold nonsymmorphic symmetries predicts 8-band Dirac features even when the platycosm formulation of the WPVZ bound fails.  Space group 73, due to the limitations of the modding procedure as it relates to inversion symmetry, combined with the body-centered geometry of its underlying lattice, is incorrectly predicted to have a minimal insulating filling of $4\mathbb{Z}$ by the platycosm formulation of the WPVZ bound~\cite{WPVZ}.  In a paper released during the final stages of preparing this manuscript, WPVZ noted using a similar eigenvalue and commutation algebra consideration that this system, generated only by two-fold nonsymmorphic operations and inversion symmetry, hosts essential eight-band features in the noninteracting limit~\cite{NewAshvin}.   

\begin{figure}
\centering
\includegraphics[width=3.5in]{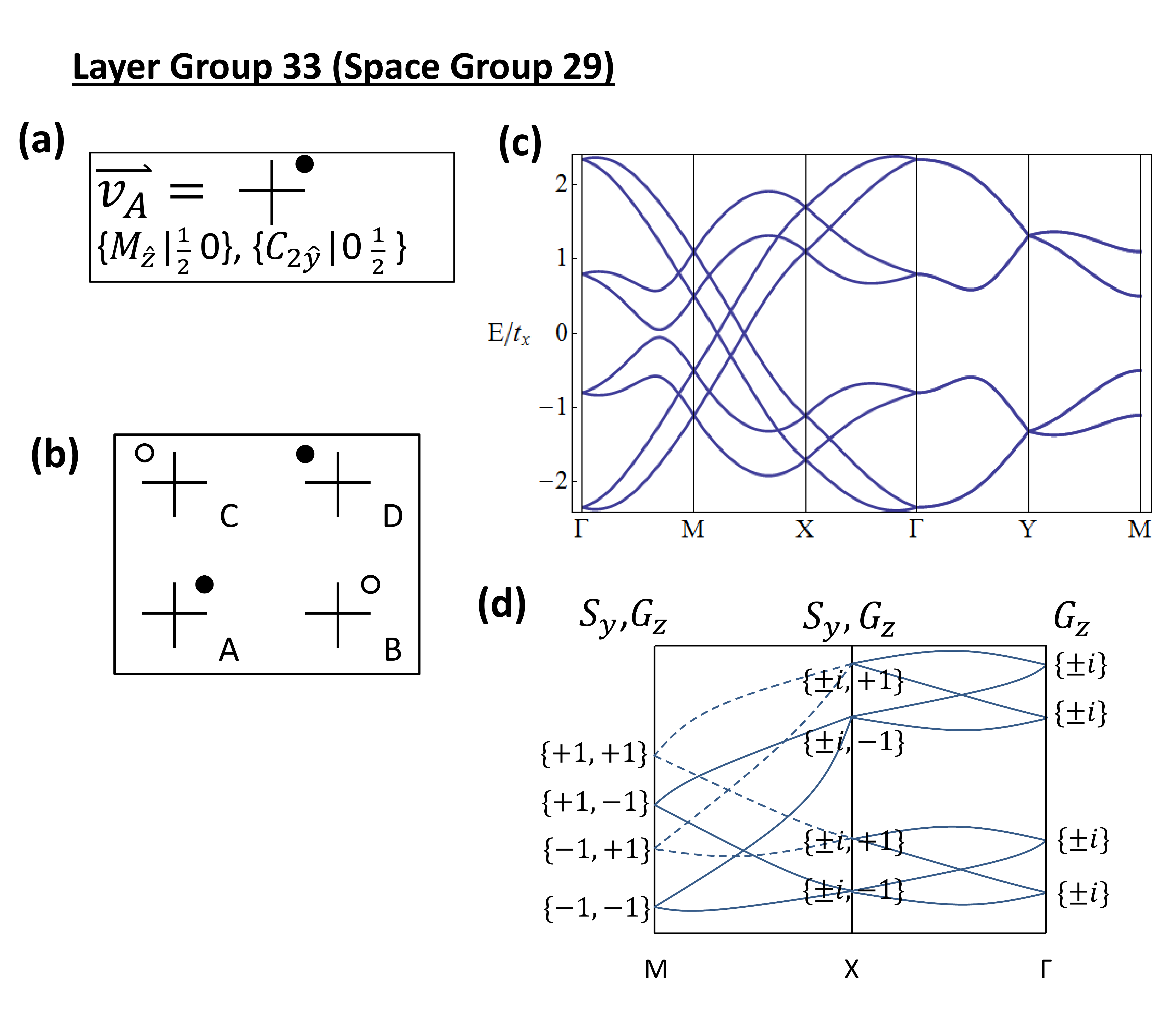}
\caption{The generators (a), lattice (b), and a typical band structure (c) for layer group 33 \emph{pb$\textit2_{1}$a} (space group 29).  This is the only layer group which  can achieve a WPVZ bound of 8 without inversion symmetry.  Bands along $\overline{\Gamma Y}$ are two-fold-degenerate by the anticommutivity of $S_{y}=t_{y/2}C_{2y}$ and $G_{x}=t_{x/2}t_{y/2}M_{\hat{x}}$.  Bands along $\overline{YM}$ are two-fold-degenerate by the combination of $G_{x}$ and $\theta$.  Even though the layer group only consists of two-fold nonsymmorphic symmetries without inversion, the combination of symmetries is such that eight bands have to be tangled together along $\overline{\Gamma X}$ and $\overline{XM}$.  Listing the eigenvalues of $S_{y}$ and $G_{z}=t_{x/2}M_{\hat{z}}$ (d), the evolution of the two-fold nonsymmorphic eigenvalues for each symmetry $\lambda_{\pm}=\pm ie^{ik_{x/y}/2}$ causes bands to form characteristic four-band structures as explained in Ref.~\onlinecite{Steve2D}.  Starting at $\Gamma$, one can choose parameters such that along $\overline{\Gamma X}$ there is a gap at half filling with these four-band structures above and below the gap.  However, because $[S_{y},G_{z}]=0$ along $\overline{XM}$, the four-band structures which form along $\overline{XM}$ preserve the eigenvalue of $G_{z}$ ($\pm1$ indicated as a dashed or solid line respectively) and exchange new partners with local protection, forming a sort of 8-band ``cat's cradle'' structure and filling in the gap at $\nu=4$ with essential Weyl points.  Should one tune parameters as to open up a gap along $\overline{XM}$, the resultant Weyl points at half filling will instead form along $\overline{\Gamma X}$.}  
\label{FigLG33}
\end{figure}

The WPVZ bound proves most advantageous when considering systems without inversion, In layer group 33 \emph{pb$\textit2_{1}$a} (space group 29) (Fig.~\ref{FigLG33}(a)), there is no inversion symmetry, and so bands can only become two-fold-degenerate by the combination of $\theta$ and a two-fold nonsymmorphic symmetry (here $G_{x}=t_{x/2}t_{y/2}M_{\hat{x}}$ for $\overline{YM}$) or by the anticommutivity of two spatial symmetries along a common line (here $S_{y}=t_{y/2}C_{2y}$ and $G_{x}$ along $\overline{\Gamma Y}$) (Fig.~\ref{FigLG33}(c)).  However, neither set of two-fold-degenerate bands contributes to obvious eight-band essential Dirac features, and one might be tempted when just considering symmetry eigenvalues to guess that this layer group is an insulator at $\nu=4$.  However, because this four-site system has $S_{y}$ and $G_{z}=t_{x/2}M_{\hat{z}}$, placement is still allowed onto the 1st amphidicosm and WPVZ predict that 8 bands must be tangled together across the 2D BZ.  Observing the band structure (Fig.~\ref{FigLG33}(c)), there is in fact an \emph{essential eight-band, ``cat's cradle-like'' Weyl feature} present along the path $\overline{\Gamma X}\cup\overline{XM}$, with all four essential Weyl points laying along the same line.

These essential Weyl points can be explained by examining the evolution of nonsymmorphic symmetry eigenvalues $\lambda_{\pm}=\pm ie^{ik_{x/y}/2}$ (Fig.~\ref{FigLG33}(d)).  Consider choosing parameters such that along $\overline{\Gamma X}$ bands are separated into two four-band, hourglass-like structures with a gap at $\nu=4$.  For most layer groups, these four-band structures would be the extent of the essential band-tangling features and the system could remain gapped at $\nu=4$ across the entire 2D BZ.  However, because in layer group 33 $[S_{y},G_{z}]=0$ along $\overline{XM}$, the four-band structures which form along $\overline{XM}$ preserve the eigenvalue of $G_{z}$ ($\pm1$ indicated as a dashed or solid line respectively in Fig.~\ref{FigLG33}(d)) and exchange new partners with local protection, forming a sort of 8-band ``cat's cradle'' structure and filling in the gap at $\nu=4$ with essential Weyl points.  Should one tune parameters as to open up a gap along $\overline{XM}$, the resultant Weyl points at half filling will instead form along $\overline{\Gamma X}$.

\section{Discussion}

In this paper, we have fully characterized the essential nodal semimetallic band features allowed in the layer groups.  By using a bound on the minimal insulating filling, derived from the compatibility between symmetry generators of a layer group and embedding the underlying lattice onto a flat compact manifold, we found that there can be layer group semimetals with interlocking groups of 2, 4, or 8 bands which cannot be untangled without lowering the spatial symmetry of the system.  This bound was achieved following a procedure by Watanabe, Po, Vishwanath, and Zaletel (WPVZ) in Ref.~\onlinecite{WPVZ} valid for both interacting and noninteracting systems which, though failing in select cases in three dimensions~\cite{DDP,NewFermions,NewAshvin}, is complete for all of the space groups which derive from trivial stackings of the wallpaper and layer groups.  Within layer groups with minimal insulating fillings of $4\mathbb{Z}$, the results of Ref.~\onlinecite{Steve2D} can be recovered, but one can also find new features, such as a band-inversion-type Dirac semimetal protected by an inversion-center offset.  Three layer group systems, specifically layer groups 33, 43, and 45 (space groups 29, 54, and 57, respectively, when stacked) can achieve minimal insulating fillings of $8\mathbb{Z}$.  Layer groups 54 and 57 have inversion symmetry, and are therefore Dirac semimetals at fillings $\nu\in4+8\mathbb{Z}$, with their 8-band essential Dirac features owing to the same mechanism of symmetry protection as the line node in SrIrO$_3$~\cite{KeeRing}.  Layer group 33, however, does not have inversion, and instead has a previously uncharacterized essential eight-band ``cat's cradle'' Weyl fermion feature with four essential Weyl points present along a high-symmetry line at fillings $\nu\in4+8\mathbb{Z}$.  

In addition to the constraints imposed on the band features of quasi-two-dimensional mono- or few-layer systems, this consideration of compact flat manifold placements, specifically for the strictly two-dimensional wallpaper systems in~\ref{sec:polygons}, also provides restrictions on the allowed band features on the surfaces of three-dimensional systems.  For groupings of bands which don't require a bulk to exist, namely the trivially-connected states of topological crystalline insulators, such as the ``hourglass fermions'' in Ref.~\onlinecite{Hourglass}, this bound indicates that \emph{symmetry can force, at most, four bands to be tangled together on the surface of a three-dimensional system}.  As there are only four wallpaper groups with glide lines, this further constrains the possible topological surface band flows as well.  Considering the allowed band features in the wallpaper groups, a combination of symmetry analysis and minimal insulating filling restrictions should allow one to exhaustively deduce all possible ``hourglass''-like surface flows permitted in bulk-insulating systems.

Finally, the two-dimensional and quasi-two-dimensional systems characterized in this paper can provide significant benefits over their three-dimensional counterparts.  They are considerably easier to visualize and analyze by crystalline symmetry.  They are also easier to simulate in tight-binding and density functional theory calculations, allowing for a relatively fast route towards predicting and engineering two-dimensional nodal semimetals, including eight-band structures analogous to those in three dimensions.  These systems can also allow experimental access to two-dimensional topological physics.  As characterized in Ref.~\onlinecite{Steve2D}, nonsymmorphic two-dimensional materials can be pinned by an additional symmetry to the quantum critical point between a trivial and a topological insulator, and therefore one could consider them as parent materials for examining strain-engineered topological phase transitions.     

We thank Toen Castle, Randall Kamien, Youngkuk Kim, and Andrew M. Rappe  for helpful discussions. This work was supported by NSF grant DMR 1120901 and a Simons Investigator grant to CLK from the Simons Foundation. 

\begin{appendix}

\section{Further Notes on and Examples of Decimations and Flat-Manifold Placement}
\label{appendix:manifolds} 

In this appendix, we visually detail the decimation procedure from Figure~\ref{FigSSH} in layered two-dimensional systems.  

\begin{figure}
\centering
\includegraphics[width=3.5in]{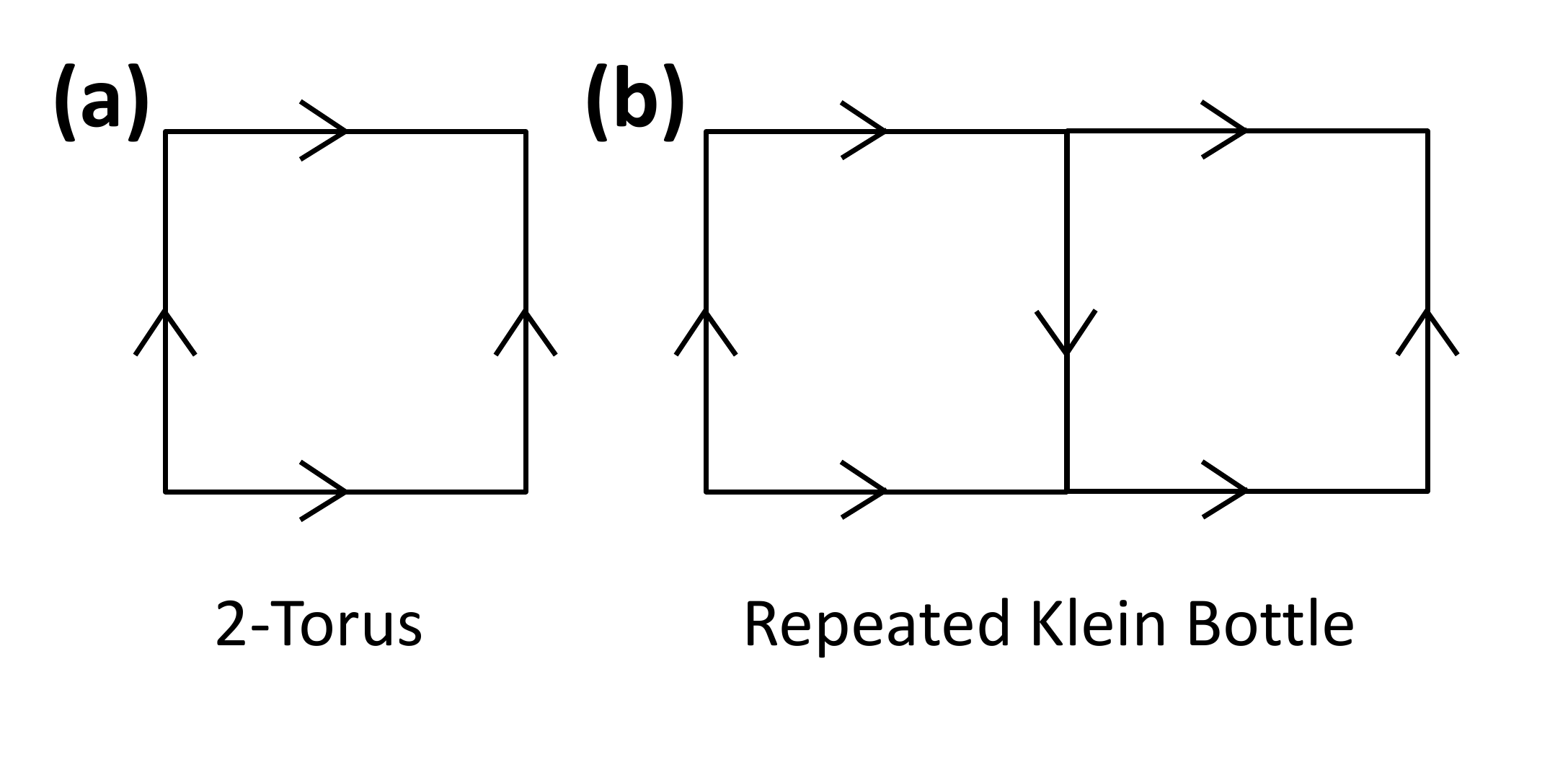}
\caption{A 2-torus (a) and two Klein bottles with a common boundary (b).  In order to create a shape with the same external arrows as the 2-torus, two Klein bottles have to be placed together, sharing the common twisted boundary.  This procedure is the origin of the bold numbers in Figures~\ref{FigWallPoly} and~\ref{FigLayerPoly}.  For a four-site unit cell, this decimation factor, $n_{dec}=A_{unit}/A_{dec}$ where $A$ is the area of the original and decimated unit cells respectively, gives the minimal insulating filling constraint $\nu\in2n_{dec}\mathbb{Z}$.}  
\label{FigDecWallpaper}
\end{figure}

For layered two-dimensional systems, the consideration of minimal insulating filling is completely captured by the number of fixed-point-free decimations of a four-site unit cell.  One can consider this unit cell as being the final one before the boundary in both the $x$ and $y$ (in-plane) directions.  As one row of atoms is chopped off and the coordinate-axis boundary condition twisted, this unit cell is decimated by modding out the nonsymmorphic symmetry which related the atoms remaining to those removed by decimation.  We can consider for any manifold a decimation factor:

\begin{equation}
n_{dec}=\frac{A_{unit}}{A_{dec}}
\end{equation}

which measures the ratio of the areas of the original to the decimated unit cell.  This factor is precisely the bold numbers indicated in Figures~\ref{FigWallPoly} and~\ref{FigLayerPoly}, which could also be expressed as the number of times a manifold would have to be repeated with a common boundary in order to create a supercell with the same external boundary as the 2-torus in wallpaper systems, or the 3-torus in general layer group systems (Figure~\ref{FigDecWallpaper}).  For a four-site unit cell in two dimensions, the insulating fillings, absent any additional band inversions, are therefore:

\begin{equation}
\nu\in2n_{dec}\mathbb{Z}.
\end{equation}

\begin{figure}
\centering
\includegraphics[width=3.5in]{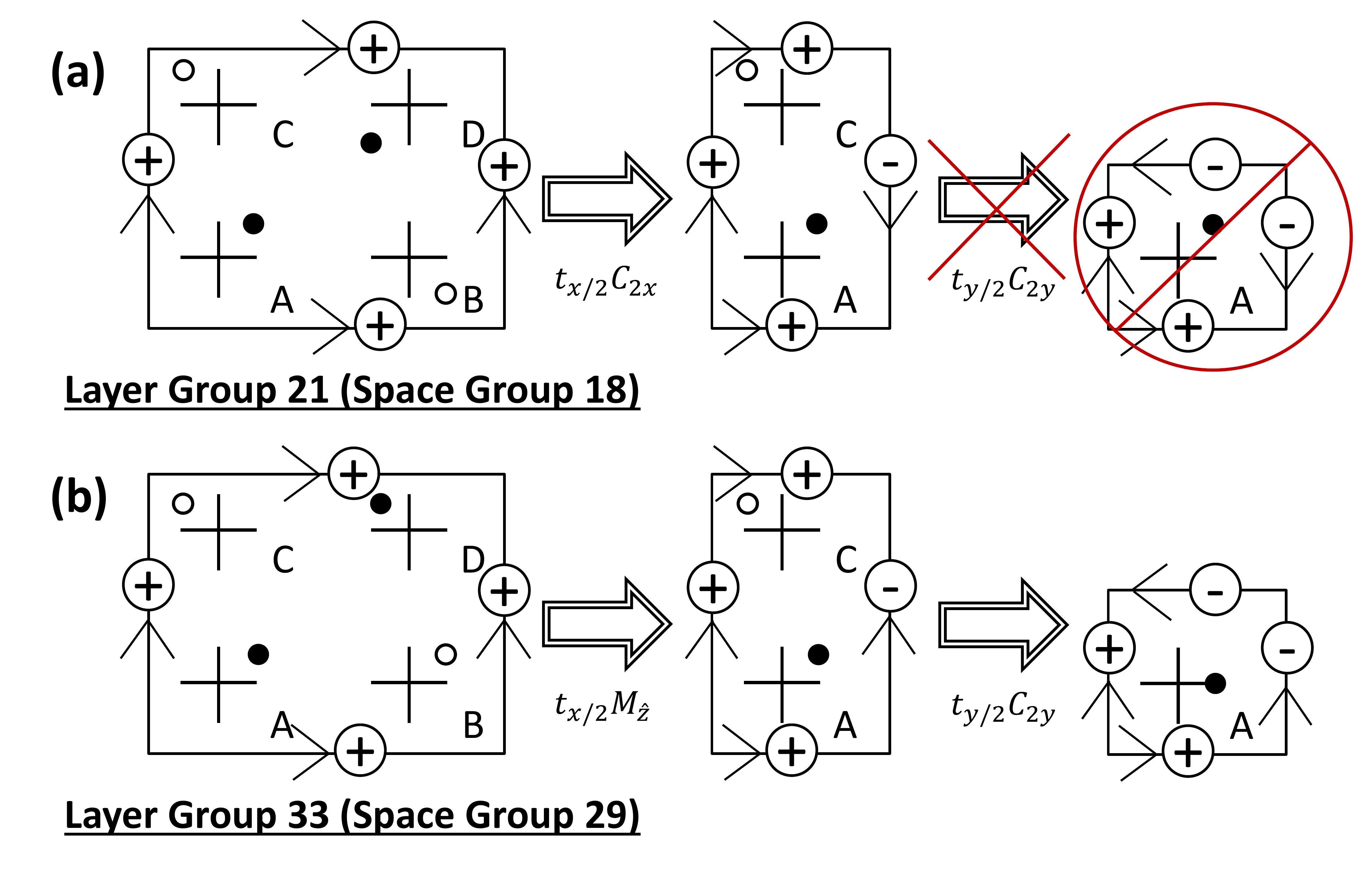}
\caption{A demonstration of the decimation procedure for two layer groups with multiple nonsymmorphic symmetries.  The minimal insulating filling is proportional to the ratio of the sizes of the maximally decimated unit cell to that of the original, with special consideration given to avoid decimations which introduce fixed points.  Layer group 21 (space group 18) (a) is generated by two perpendicular screws: $S_{x/y}=t_{x/y/2}C_{2x/y}$.  One could choose to mod out $S_{x}$ first, reducing the area of the unit cell by half and placing the system onto the dicosm.  However, further decimation by $S_{y}$ would then be disallowed, because $S_{x}S_{y}\sim t_{x/2}t_{y/2}(C_{2z})$, which is an inherently \emph{symmorphic} operation ($C_{2Z}$ about the center of the unit cell).  Therefore, choosing either screw, the maximal decimation of layer group 21 gives $n_{dec}=2$ and placement onto the dicosm, with a minimal insulating filling of $\nu\in4\mathbb{Z}$.  Conversely, layer group 33 (space group 29) (b) is generated by $S_{y}$ and $G_{z}=t_{x/2}M_{\hat{z}}$.  Modding out $G_{z}$ first removes the right half of the unit cell and places the system onto the 1st amphicosm.  However, this is not the maximal decimation, as the product $S_{y}G_{z}\sim t_{x/2}(t_{y/2}M_{\hat{x}})$, which is an inherently \emph{nonsymmorphic} operation.  Therefore, layer group 33 admits an additional decimation by $S_{y}$ onto the 1st amphidicosm, resulting in $n_{dec}=4$ and a minimal insulating filling of $\nu\in8\mathbb{z}$.}  
\label{FigDecLayer}
\end{figure}

A central part of this procedure is the restriction that we only utilized fixed-point-free decimations.  As emphasized in the main text, \emph{multiple decimations by two-fold operations are only allowed if the product of those operations is itself also an inherently nonsymmorphic operation}~\cite{WPVZ,Conway}.  

Figure~\ref{FigDecLayer} illustrates two examples of layer groups with multiple nonsymmorphic group elements.  Layer group 21 (space group 18) (Fig.~\ref{FigDecLayer}(a)) is generated by two perpendicular screws: $S_{x/y}=t_{x/y/2}C_{2x/y}$.  One could choose to mod out $S_{x}$ first, reducing the area of the unit cell by half and placing the system onto the dicosm.  However, further decimation by $S_{y}$ would then be disallowed, because $S_{x}S_{y}\sim t_{x/2}t_{y/2}(C_{2z})$, which is an inherently \emph{symmorphic} operation ($C_{2Z}$ about the center of the unit cell).  Therefore, choosing either screw, the maximal decimation of layer group 21 gives $n_{dec}=2$ and placement onto the dicosm, with a minimal insulating filling of $\nu\in4\mathbb{Z}$.  Conversely, layer group 33 (space group 29) (Fig.~\ref{FigDecLayer}(b)) is generated by $S_{y}$ and $G_{z}=t_{x/2}M_{\hat{z}}$.  Modding out $G_{z}$ first removes the right half of the unit cell and places the system onto the 1st amphicosm.  However, this is not the maximal decimation, as the product $S_{y}G_{z}\sim t_{x/2}(t_{y/2}M_{\hat{x}})$, which is an inherently \emph{nonsymmorphic} operation.  Therefore, layer group 33 admits an additional decimation by $S_{y}$ onto the 1st amphidicosm, resulting in $n_{dec}=4$ and a minimal insulating filling of $\nu\in8\mathbb{Z}$.

\section{Tight-Binding Models} 
\label{appendix:tb}

In this appendix, we list the tight-binding Hamiltonians for the example layer groups selected for this manuscript (Figs.~\ref{FigLG37}-\ref{FigLG33}).  

We begin by considering a four-site rectangular unit cell in two dimensions (Fig.~\ref{FigLattice}).  Initially, each site has spherical symmetry such that this high-symmetry system can be considered a relabeling of a rectangular one-site unit cell with all of the symmetries of the underlying Bravais lattice.  Repeating the procedure from~\ref{sec:LGs}, we designate Pauli matrices for the sublattice degrees of freedom, with $\tau^{x}$ indicating s-orbital-like hopping between the A and B (and C and D) sites and $\mu^{x}$ indicating s-orbital-like hopping between the A and C sites (and B and D) such that $\tau^{x}\mu^{x}$ indicates s-like-hopping between A and D sites (and B and C).  Each site is given an additional spin degree of freedom $\sigma$ such that our overall model has eight bands.  

We can, in this high-symmetry limit, first write down all of the first- and second-nearest-neighbor s-orbital-like hopping terms:

\begin{eqnarray}
\mathcal{H}_{0} &=& t_{x}\cos\left(\frac{k_{x}}{2}\right)\tau^{x} + t_{y}\cos\left(\frac{k_{y}}{2}\right)\mu^{x} \nonumber \\
&+& t_{2}\cos\left(\frac{k_{x}}{2}\right)\cos\left(\frac{k_{y}}{2}\right)\tau^{x}\mu^{x}
\end{eqnarray}

where the lattice spacing $a_{x/y}$ has been set to 1.  From there, terms can be added to reduce the symmetry of the system into a particular layer group and lift many of the degeneracies.  For a particular layer group $LG$, we consider the set of all symmetry-allowed first- and second-nearest-neighbor hoppings (other than the existing s-like ones) to be a potential $V_{LG}$ such that overall $\mathcal{H}_{LG}=\mathcal{H}_{0} + V_{LG}$.  We find that using all of the terms up to second-nearest-neighbor interactions produces layer-group-specific band structures (though occasionally allows for an artificial chiral symmetry, which could be broken by introducing symmetry-allowed third-nearest-neighbor terms).   

For each layer group, the allowed terms $V_{LG}$ can be determined by considering how the group generators transform a generic Hamiltonian $\mathcal{H}(k_{x},k_{y})$ and finding all physical hopping terms invariant under that transformation.  In practice, the form of the representation of each group generator in the sublattice and Bloch space is that of the representation at $\Gamma$ multiplied by an operation on $\vec{k}$.  Additionally, all systems are considered to be time-reversal symmetric with $\theta=i\sigma^{y}K\otimes(\vec{k}\rightarrow -\vec{k})$ such that $\theta^{2}=-1$.    

In the following subsections, we detail the group generators and $V_{LG}$ for each of the example systems in the text.  All generators are defined from the common origin of site A.  In order, the generators are described by the name of the generator, the generator as operations defined from site A, and the form of the representation of the operator for our eight-band $\mathcal{H}(\vec{k})$.  Similar terms are grouped under the same constants for simplicity, though this is not explicity required by symmetry.  Values of the constants used for band plots have been noted after each $V_{LG}$. 

\subsection{Layer Group 37}  

Layer group 37, \emph{pmmm} (space group 47) has a WPVZ bound of 2 and the following generators:

\begin{eqnarray}
M_{\hat{x}} = t_{x/2}M_{\hat{x}} &=& \tau^{x}\sigma^{x}\otimes(k_{x}\rightarrow -k_{x}) \nonumber \\
M_{\hat{y}} = t_{y/2}M_{\hat{y}} &=& \mu^{x}\sigma^{y}\otimes(k_{y}\rightarrow -k_{y}) \nonumber \\
M_{\hat{z}} = M_{\hat{z}} &=& \sigma^{z}.
\end{eqnarray}

This results in the following allowed first- and second-nearest-neighbor hopping terms:

\begin{eqnarray}
V_{37} &=& \cos\left(\frac{k_{x}}{2}\right)\left[v_{r1x}\tau^{y}\mu^{z}\sigma^{z}\right] \nonumber \\
&+& \sin\left(\frac{k_{x}}{2}\right)\left[v_{p1x}\tau^{y} + v_{s1x}\tau^{x}\mu^{z}\sigma^{z}\right] \nonumber \\
&+& \cos\left(\frac{k_{y}}{2}\right)\left[v_{r1y}\tau^{z}\mu^{y}\sigma^{z}\right] \nonumber \\
&+& \sin\left(\frac{k_{y}}{2}\right)\left[v_{p1y}\mu^{y} + v_{s1y}\tau^{z}\mu^{x}\sigma^{z}\right] \nonumber \\
&+& \sin\left(\frac{k_{x}}{2}\right)\cos\left(\frac{k_{y}}{2}\right)\left[v_{p2}\tau^{y}\mu^{x}\right] \nonumber \\
&+& \cos\left(\frac{k_{x}}{2}\right)\sin\left(\frac{k_{y}}{2}\right)\left[v_{p2}\tau^{x}\mu^{y}\right] \nonumber \\
&+& \sin\left(\frac{k_{x}}{2}\right)\sin\left(\frac{k_{y}}{2}\right)\left[v_{p2}\tau^{y}\mu^{y}\right].
\label{V37}
\end{eqnarray}

For the bands in Fig.~\ref{FigLG37}(c),

\begin{eqnarray}
t_{x} &=& 1.0,\ t_{y}=1.25,\ t_{2}=0.4,\ v_{r1x}=0.3, \nonumber \\
v_{p1x} &=& 0.35,\ v_{s1x}=-0.65,\ v_{r1y}=0.45,\ v_{p1y}=0.65 \nonumber \\
v_{vs1y} &=& -0.8,\ v_{p2}=-0.2.
\end{eqnarray}

\subsection{Layer Group 23}

Layer group 23, \emph{pmm2} (space group 25), has a WPVZ bound of 2.  It is the result of breaking $M_{\hat{z}}$ in layer group 37, leading to the following generators:

\begin{eqnarray}
M_{\hat{x}} = t_{x/2}M_{\hat{x}} &=& \tau^{x}\sigma^{x}\otimes(k_{x}\rightarrow -k_{x}) \nonumber \\
M_{\hat{y}} = t_{y/2}M_{\hat{y}} &=& \mu^{x}\sigma^{y}\otimes(k_{y}\rightarrow -k_{y}).
\end{eqnarray}

The allowed terms in layer group 23 can be therefore considered as those allowed for layer group 37, plus new terms which don't commute with $\sigma^{z}$:

\begin{eqnarray}
V_{23} &=& V_{37} \nonumber \\
&+& \cos\left(\frac{k_{x}}{2}\right)\left[v_{r1x}\tau^{y}\sigma^{y}\right] + \sin\left(\frac{k_{x}}{2}\right)\left[v_{s1x}\tau^{x}\sigma^{y}\right] \nonumber \\
&+& \cos\left(\frac{k_{y}}{2}\right)\left[v_{r1y}\mu^{y}\sigma^{x}\right] + \sin\left(\frac{k_{y}}{2}\right)\left[v_{s1y}\mu^{x}\sigma^{x}\right] \nonumber \\
&+& \cos\left(\frac{k_{x}}{2}\right)\cos\left(\frac{k_{y}}{2}\right)\left[v_{r2}\left(\tau^{x}\mu^{y}\sigma^{x} + \tau^{y}\mu^{x}\sigma^{y}\right)\right] \nonumber \\
&+& \sin\left(\frac{k_{x}}{2}\right)\cos\left(\frac{k_{y}}{2}\right)\left[v_{s2}\left(\tau^{x}\mu^{x}\sigma^{y} + \tau^{y}\mu^{y}\sigma^{x}\right)\right] \nonumber \\
&+& \cos\left(\frac{k_{x}}{2}\right)\sin\left(\frac{k_{y}}{2}\right)\left[v_{s2}\left(\tau^{x}\mu^{x}\sigma^{x} + \tau^{y}\mu^{y}\sigma^{y}\right)\right] \nonumber \\
&+& \sin\left(\frac{k_{x}}{2}\right)\sin\left(\frac{k_{y}}{2}\right)\left[v_{r2}\left(\tau^{x}\mu^{y}\sigma^{y} + \tau^{y}\mu^{x}\sigma^{x}\right)\right]. \nonumber \\
\end{eqnarray}

For the bands in Fig.~\ref{FigLG23}(c),

\begin{eqnarray}
t_{x} &=& 1.0,\ t_{y}=1.25,\ t_{2}=0.1,\ v_{r1x}=0.3, \nonumber \\
v_{p1x} &=& 0.35,\ v_{s1x}=-0.65,\ v_{r1y}=0.45,\ v_{p1y}=0.65 \nonumber \\
v_{vs1y} &=& 0.7,\ v_{p2}=-0.2,\ v_{s2}=-0.35,\ v_{r2}=0.3.
\end{eqnarray}

For the bands in Fig.~\ref{FigLG23}(d),

\begin{eqnarray}
t_{x} &=& 1.0,\ t_{y}=1.25,\ t_{2}=0.9,\ v_{r1x}=0.3, \nonumber \\
v_{p1x} &=& 0.35,\ v_{s1x}=-0.65,\ v_{r1y}=0.45,\ v_{p1y}=0.65 \nonumber \\
v_{vs1y} &=& 0.7,\ v_{p2}=-0.9,\ v_{s2}=-0.35,\ v_{r2}=0.1.
\end{eqnarray}

\subsection{Layer Group 44}

Layer group 44, \emph{pbam} (space group 55) has a WPVZ bound of 4 and the following generators:

\begin{eqnarray}
S_{x} = t_{x/2}C_{2x} &=& \tau^{x}\sigma^{x}\otimes(k_{y}\rightarrow -k_{y}) \nonumber \\
S_{y} = t_{y/2}C_{2y} &=& \mu^{x}\sigma^{y}\otimes(k_{x}\rightarrow -k_{x}) \nonumber \\
M_{\hat{z}} = M_{\hat{z}} = \sigma^{z}. 
\end{eqnarray}

This results in the following allowed first- and second-nearest-neighbor hopping terms:

\begin{eqnarray}
V_{44} &=& \cos\left(\frac{k_{x}}{2}\right)\left[v_{r1x}\tau^{y}\mu^{z}\sigma^{z}\right] + \cos\left(\frac{k_{y}}{2}\right)\left[v_{r1y}\tau^{z}\mu^{y}\sigma^{z}\right] \nonumber \\
&+& \sin\left(\frac{k_{x}}{2}\right)\cos\left(\frac{k_{y}}{2}\right)\left[v_{p2}\tau^{x}\mu^{y}\right] \nonumber \\
&+& \cos\left(\frac{k_{x}}{2}\right)\sin\left(\frac{k_{y}}{2}\right)\left[v_{p2}\tau^{y}\mu^{x}\right] \nonumber \\
&+& \sin\left(\frac{k_{x}}{2}\right)\sin\left(\frac{k_{y}}{2}\right)\left[v_{p2}\tau^{y}\mu^{y}\right].
\end{eqnarray}

For the bands in Fig.~\ref{FigLG44}(c),  

\begin{eqnarray}
t_{x} &=& 1.0,\ t_{y}=1.55,\ t_{2}=0.4,\ v_{r1x}=0.3, \nonumber \\
v_{vr1y} &=& 0.6,\ v_{p2}=0.2.
\end{eqnarray}

\subsection{Layer Group 21}

Layer group 21, \emph{p$\textit2_{1}\textit2_{1}$2} (space group 18), has a WPVZ bound of 4.  It is the result of breaking $M_{\hat{z}}$ in layer group 44, leading to the following generators:

\begin{eqnarray}
S_{x} = t_{x/2}C_{2x} &=& \tau^{x}\sigma^{x}\otimes(k_{y}\rightarrow -k_{y}) \nonumber \\
S_{y} = t_{y/2}C_{2y} &=& \mu^{x}\sigma^{y}\otimes(k_{x}\rightarrow -k_{x}). 
\end{eqnarray}

The allowed terms in layer group 21 can be therefore considered as those allowed for layer group 44, plus new terms which don't commute with $\sigma^{z}$:

\begin{eqnarray}
V_{21} &=& V_{44} \nonumber \\
&+& \cos\left(\frac{k_{x}}{2}\right)\left[v_{r1x}\left(\tau^{y}\sigma^{y} + \tau^{y}\mu^{z}\sigma^{z}\right)\right] \nonumber \\
&+& \sin\left(\frac{k_{x}}{2}\right)\left[v_{s1x}\tau^{x}\sigma^{x}\right] \nonumber \\
&+& \cos\left(\frac{k_{y}}{2}\right)\left[v_{r1y}\left(\mu^{y}\sigma^{x} + \tau^{z}\mu^{y}\sigma^{z}\right)\right] \nonumber \\
&+& \sin\left(\frac{k_{y}}{2}\right)\left[v_{s1y}\mu^{x}\sigma^{y}\right] \nonumber \\
&+& \cos\left(\frac{k_{x}}{2}\right)\cos\left(\frac{k_{y}}{2}\right)\left[v_{r2}\left(\tau^{x}\mu^{y}\sigma^{x} + \tau^{y}\mu^{x}\sigma^{y}\right)\right] \nonumber \\
&+& \sin\left(\frac{k_{x}}{2}\right)\cos\left(\frac{k_{y}}{2}\right)\left[v_{s2}\left(\tau^{x}\mu^{x}\sigma^{x} + \tau^{y}\mu^{y}\sigma^{y}\right)\right] \nonumber \\
&+& \cos\left(\frac{k_{x}}{2}\right)\sin\left(\frac{k_{y}}{2}\right)\left[v_{s2}\left(\tau^{x}\mu^{x}\sigma^{y} + \tau^{y}\mu^{y}\sigma^{x}\right)\right] \nonumber \\
&+& \sin\left(\frac{k_{x}}{2}\right)\sin\left(\frac{k_{y}}{2}\right)\left[v_{r2}\left(\tau^{x}\mu^{y}\sigma^{y} + \tau^{y}\mu^{x}\sigma^{x}\right)\right]. \nonumber \\
\end{eqnarray}

For the bands in Fig.~\ref{FigLG21}(c),

\begin{eqnarray}
t_{x} &=& 1.0,\ t_{y}=1.55,\ t_{2}=0.4,\ v_{r1x}=0.3, \nonumber \\
v_{s1x} &=& 0.65,\ v_{vr1y} = 0.6,\ v_{s1y}=0.85, \nonumber \\
v_{r2} &=& 0.6,\ v_{s2}= 0.7. 
\end{eqnarray}

\subsection{Layer Group 17}

Layer group 17, \emph{p$\textit2_{1}/b$11} (space group 14) has a WPVZ bound of 4 and the following generators:

\begin{eqnarray}
P = t_{y/2}P &=& \mu^{x}\otimes(\vec{k}\rightarrow -\vec{k}) \nonumber \\
G_{y} = t_{x/2}M_{\hat{y}} &=& \tau^{x}\sigma^{y}\otimes(k_{y}\rightarrow -k_{y}). 
\end{eqnarray}

This results in the following allowed first- and second-nearest-neighbor hopping terms:

\begin{eqnarray}
V_{17} &=& \cos\left(\frac{k_{x}}{2}\right)\left[v_{r1x}\tau^{y}\sigma^{x}\right] + \sin\left(\frac{k_{x}}{2}\right)\left[v_{s1x}\tau^{x}\mu^{z}\sigma^{y}\right] \nonumber \\
&+& \sin\left(\frac{k_{y}}{2}\right)\left[v_{p1y}\tau^{z}\mu^{y}\right] \nonumber \\
&+& \cos\left(\frac{k_{x}}{2}\right)\cos\left(\frac{k_{y}}{2}\right)\left[v_{r2}\tau^{y}\mu^{x}\sigma^{x}\right] \nonumber \\
&+& \sin\left(\frac{k_{x}}{2}\right)\cos\left(\frac{k_{y}}{2}\right)\left[v_{p2}\tau^{x}\mu^{y} + v_{s2}\tau^{y}\mu^{y}\sigma^{x}\right] \nonumber \\
&+& \cos\left(\frac{k_{x}}{2}\right)\sin\left(\frac{k_{y}}{2}\right)\left[v_{s2}\tau^{y}\mu^{y}\sigma^{y}\right] \nonumber \\
&+& \sin\left(\frac{k_{x}}{2}\right)\sin\left(\frac{k_{y}}{2}\right)\left[v_{r2}\tau^{y}\mu^{x}\sigma^{y}\right]
\end{eqnarray}

noting that additional terms are also allowed due to this system's invariance under $\sigma^{x}\leftrightarrow\sigma^{z}$.  

For the bands in Fig.~\ref{FigLG17}(c),

\begin{eqnarray}
t_{x} &=& 1.0,\ t_{y}=1.14,\ t_{2}=0.4,\ v_{r1x}=0.3, \nonumber \\
v_{s1x} &=& 0.65,\ v_{vp1y} = 0.8,\ v_{r2}=0.25, \nonumber \\ 
v_{p2}&=&0.2,\ v_{s2}=0.45. 
\end{eqnarray}

For the bands in Fig.~\ref{FigLG17}(d),

\begin{eqnarray}
t_{x} &=& 1.0,\ t_{y}=1.14,\ t_{2}=0.4,\ v_{r1x}=0.3, \nonumber \\
v_{s1x} &=& 0.65,\ v_{vp1y} = 0.2,\ v_{r2}=0.25, \nonumber \\ 
v_{p2}&=&0.2,\ v_{s2}=0.45. 
\end{eqnarray}

\subsection{Layer Group 45}

Layer group 45, \emph{pbma} (space group 57) has a WPVZ bound of 8 and the following generators:

\begin{eqnarray}
G_{z} = t_{x/2}M_{\hat{z}} &=& \tau^{x}\sigma^{z} \nonumber \\
S_{y} = t_{x/2}t_{y/2}C_{2y} &=& \tau^{x}\mu^{x}\sigma^{y}\otimes(k_{x}\rightarrow -k_{x}) \nonumber \\
P = t_{x/2}t_{y/2}P &=& \tau^{x}\mu^{x}\otimes(\vec{k}\rightarrow -\vec{k}).
\end{eqnarray}

This results in the following allowed first- and second-nearest-neighbor hopping terms:

\begin{eqnarray}
V_{45} &=& \cos\left(\frac{k_{x}}{2}\right)\left[v_{r1x}\tau^{y}\mu^{z}\sigma^{y}\right] + \cos\left(\frac{k_{y}}{2}\right)\left[v_{r1y}\tau^{z}\mu^{y}\sigma^{y}\right] \nonumber \\
&+& \sin\left(\frac{k_{y}}{2}\right)\left[v_{s1y}\tau^{z}\mu^{x}\sigma^{x}\right] \nonumber \\
&+& \sin\left(\frac{k_{x}}{2}\right)\cos\left(\frac{k_{y}}{2}\right)\left[v_{p2}\tau^{x}\mu^{y}\right]. 
\end{eqnarray}

For the bands in Fig.~\ref{FigLG45}(c),

\begin{eqnarray}
t_{x} &=& 1.0,\ t_{y}=1.25,\ t_{2}=0.4,\ v_{r1x}=-0.3, \nonumber \\
v_{r1y} &=& 0.45,\ v_{vs1y} = 0.8,\ v_{p2}=-0.2.
\end{eqnarray}

\subsection{Layer Group 33}

Layer group 33, \emph{pb$\textit2_{1}$a} (space group 29) has a WPVZ bound of 8 and the following generators:

\begin{eqnarray}
G_{z} = t_{x/2}M_{\hat{z}} &=& \tau^{x}\sigma^{z} \nonumber \\
S_{y} = t_{y/2}C_{2y} &=& \mu^{x}\sigma^{y}\otimes(k_{x}\rightarrow -k_{x}).
\end{eqnarray}

\begin{eqnarray}
V_{33} &=& \cos\left(\frac{k_{x}}{2}\right)\left[v_{r1x}\tau^{y}\mu^{z}\sigma^{x}\right] + \sin\left(\frac{k_{x}}{2}\right)\left[v_{s1x}\tau^{x}\sigma^{z}\right] \nonumber \\
&+& \cos\left(\frac{k_{y}}{2}\right)\left[v_{r1y}\left(\tau^{z}\mu^{y}\sigma^{x} + \mu^{y}\sigma^{z}\right)\right] \nonumber \\
&+& \sin\left(\frac{k_{y}}{2}\right)\left[v_{s1y}\tau^{z}\mu^{x}\sigma^{y}\right] \nonumber \\
&+& \cos\left(\frac{k_{x}}{2}\right)\cos\left(\frac{k_{y}}{2}\right)\left[v_{r2}\left(\tau^{y}\mu^{x}\sigma^{y} + \tau^{x}\mu^{y}\sigma^{z}\right)\right] \nonumber \\
&+& \sin\left(\frac{k_{x}}{2}\right)\cos\left(\frac{k_{y}}{2}\right)\ \times \nonumber \\
& &\left[v_{p2}\tau^{x}\mu^{y} + v_{s2}\left(\tau^{y}\mu^{y}\sigma^{y} + \tau^{x}\mu^{x}\sigma^{z}\right)\right] \nonumber \\
&+& \cos\left(\frac{k_{x}}{2}\right)\sin\left(\frac{k_{y}}{2}\right)\left[v_{s2}\tau^{y}\mu^{y}\sigma^{x}\right] \nonumber \\
&+& \sin\left(\frac{k_{x}}{2}\right)\sin\left(\frac{k_{y}}{2}\right)\left[v_{r2}\tau^{y}\mu^{x}\sigma^{x}\right]. 
\end{eqnarray}

For the bands in Fig.~\ref{FigLG33}(c),

\begin{eqnarray}
t_{x} &=& 1.0,\ t_{y}=1.25,\ t_{2}=0.4,\ v_{r1x}=-0.3, \nonumber \\
v_{s1x} &=& 0.3,\ v_{r1y} = 0.45,\ v_{vs1y} = 0.8, \nonumber \\
v_{r2} &=& 0.25,\ v_{p2} = -0.2,\ v_{s2} = 0.45.
\end{eqnarray}

\section{List of Filling Conditions for the 80 Layer Groups}
\label{appendix:fillingcond} 

In this appendix, we list the 80 layer groups as sorted by their allowed platycosm placements.  For each layer group, we cite the equivalent space group for a three-dimensional stack of that system~\cite{LGtoSGconvert}.  

The 17 layer groups which could additionally describe the boundary of a three-dimensional object also comprise the \emph{wallpaper groups} and are denoted with (w).  These groups contain no operations which would exchange the interior and exterior of such a three-dimensional object, and therefore (if $\hat{z}$ is the layer stacking direction or surface normal) are disallowed from having $P$, $M_{\hat{z}}$, or $C_{2x/y}$, as well as any of those operations followed by a fractional lattice translation.   

Layer groups without nonsymmorphic symmetries are only allowed placement onto the torocosm, or 3-torus, and have no insulating filling constraints besides $\nu\in2\mathbb{Z}$ by Kramers' theorem. 

\begin{center}
\begin{tabular}{ |c|c|c|c|c| } 
\hline
\multicolumn{5}{|c|}{3-Torus \emph{(43 Layer Groups)}} \\
\multicolumn{5}{|c|}{$\nu\in2\mathbb{Z}$} \\
 \hline
Layer Group & Space Group & & Layer Group & Space Group \\
\hline
 1 (w) & 1 &\ \ \ \ & 53 & 89 \\
 2 & 2 & & 55 (w) & 99 \\
 3 (w) & 3 & & 57 & 111 \\
 4 & 6 & & 59 & 115 \\
 6 & 10 & & 61 & 123 \\
 8 & 3 & & 65 (w) & 143 \\
 10 & 5 & & 66 & 147 \\
 11 (w) & 6 & & 67 & 149 \\
 13 (w) & 8 & & 68 & 150 \\
 14 & 10 & & 69 (w) & 156 \\
 18 & 12 & & 70 (w) & 157 \\
 19 & 16 & & 71 & 162 \\
 22 & 21 & & 72 & 164 \\
 23 (w) & 25 & & 73 (w) & 168 \\
 26 (w) & 35 & & 74 & 174 \\
 27 & 25 & & 75 & 175 \\
 35 & 35 & & 76 & 177 \\
 37 & 47 & & 77 (w) & 183 \\
 47 & 65 & & 78 & 187 \\
 49 (w) & 75 & & 79 & 189 \\
 50 & 81 & & 80 & 191 \\
 51 & 83 & &  &  \\
 \hline
\end{tabular}
\end{center}

Layer groups with two-fold screws and no glide mirrors can be decimated and placed onto the two-sided dicosm, which results in insulating fillings of $\nu\in4\mathbb{Z}$ absent any additional band inversions with locally-protected crossings (otherwise stated as the ``minimal-insulating filling'').  

\begin{center}
\begin{tabular}{ |c|c|c|c|c| } 
\hline
\multicolumn{5}{|c|}{Only Dicosm \emph{(6 Layer Groups)}} \\
\multicolumn{5}{|c|}{$\nu\in4\mathbb{Z}$} \\
 \hline
Layer Group & Space Group & & Layer Group & Space Group \\
\hline
 9 & 4 &\ \ \ \ & 21 & 18 \\
15  & 11 & & 54 & 90 \\
20 & 17 & & 58 & 113 \\
 \hline
\end{tabular}
\end{center}

Layer groups with glide mirrors and no two-fold screws can be decimated and placed onto the one-sided 1st amphicosm, which results in a minimal insulating filling of $\nu\in4\mathbb{Z}$.  

Layer groups with both glide mirrors and two-fold screws require more careful examination.  As part of the procedure for decimation from~\ref{sec:platycosms} and Appendix~\ref{appendix:manifolds}, all combinations of perpendicular nonsymmorphic symmetries must be examined to determine if further decimation is allowed from the dicosm or 1st amphicosm into the 1st amphidicosm.  \emph{For the layer groups, that decimation is in practice only allowed for systems with four or more sites per unit cell and any $z$-axis rotation of $S_{y}=t_{y/2}C_{2y}$ and $G_{z}=t_{x/2}M_{\hat{z}}$. }  

\begin{center}
\begin{tabular}{ |c|c|c|c|c| } 
\hline
\multicolumn{5}{|c|}{Only 1st Amphicosm \emph{(17 Layer Groups)}} \\
\multicolumn{5}{|c|}{$\nu\in4\mathbb{Z}$} \\
 \hline
Layer Group & Space Group & & Layer Group & Space Group \\
\hline
 5 & 7  &\ \ \ \ & 36 & 39 \\
7 & 13 & & 38 & 49 \\
12 (w) & 7 & & 39 & 50 \\
16 & 13 & & 48 & 67 \\
24 (w) & 28 & & 52 & 85 \\
25 (w) & 32 & & 56 (w) & 100 \\
30 & 27 & & 60 & 117 \\
31 & 28 & & 62 & 125 \\
34 & 30 & & & \\
 \hline
\end{tabular}
\end{center}

Absent these conditions, for layer groups with both glide mirrors and screws, frequently the case in those with inversion symmetry, one could choose to mod out using  \emph{either} the glide or the two-fold screw, allowing placement onto either the dicosm or the 1st amphidicosm.  For both cases, the filling restrictions are the same: an insulator can only occur at fillings of $\nu\in4\mathbb{Z}$.  

\begin{center}
\begin{tabular}{ |c|c|c|c|c| } 
\hline
\multicolumn{5}{|c|}{Dicosm or 1st Amphicosm \emph{(11 Layer Groups)}} \\
\multicolumn{5}{|c|}{$\nu\in4\mathbb{Z}$} \\
 \hline
Layer Group & Space Group & & Layer Group & Space Group \\
\hline
 17 & 14  &\ \ \ \ & 42 & 53 \\
28 & 26 & & 44 & 55 \\
29 & 26 & & 46 & 59 \\
32 & 31 & & 63 & 127 \\
40 & 51 & & 64 & 129 \\
41 & 51 & & & \\
 \hline
\end{tabular}
\end{center}

Finally, these conditions for further decimation onto to the one-sided 1st amphidicosm are, in fact, only satisfied by 3 layer groups.  For these groups, eight bands have to be tangled together, and therefore these 3 layer groups have minimal insulating fillings of $\nu\in8\mathbb{Z}$.  

\begin{center}
\begin{tabular}{ |c|c|c|c|c| } 
\hline
\multicolumn{5}{|c|}{1st Amphidicosm \emph{(3 Layer Groups)}} \\
\multicolumn{5}{|c|}{$\nu\in8\mathbb{Z}$} \\
 \hline
Layer Group & Space Group & & Layer Group & Space Group \\
\hline
 33 & 29  &\ \ \ \ & 45 & 57 \\
43 & 54 & & & \\
 \hline
\end{tabular}
\end{center}

\end{appendix}

\end{document}